\documentclass[aps,
pra,
reprint,
floatfix,
superscriptaddress,
]{revtex4-1}

\PassOptionsToPackage{hypertexnames=false}{hyperref}

\usepackage{amsmath}
\usepackage{amsfonts}
\usepackage{amssymb}
\usepackage{graphicx}
\usepackage{accents}
\usepackage[hidelinks]{hyperref}
\usepackage{enumitem}
\usepackage[section]{placeins}

\usepackage{svg}
\usepackage[caption=false]{subfig}
\usepackage{tikz} % for \filledtriangle
\usepackage{letterspace} % for SWAP
\usepackage{soul}
\usepackage{yhmath}

\usepackage{pifont} % for the big dot \ding{108}

\newcommand{\Tr}{\mbox{Tr}}

\newcommand{\ket}[1]{|#1\rangle}
\newcommand{\bra}[1]{\langle #1|}

\makeatletter % the lines below make the double bra and double ket symbols
\newsavebox{\@brx}
\newcommand{\llangle}[1][]{\savebox{\@brx}{\(\m@th{#1\langle}\)}%
  \mathopen{\copy\@brx\kern-0.5\wd\@brx\usebox{\@brx}}}
\newcommand{\rrangle}[1][]{\savebox{\@brx}{\(\m@th{#1\rangle}\)}%
  \mathclose{\copy\@brx\kern-0.5\wd\@brx\usebox{\@brx}}}
\makeatother

\newlength{\dhatheight} % lines below make the double-hat symbol

\newcommand{\qed}{\nobreak \ifvmode \relax \else
      \ifdim\lastskip<1.5em \hskip-\lastskip
      \hskip1.5em plus0em minus0.5em \fi \nobreak
      \vrule height0.75em width0.5em depth0.25em\fi}

\definecolor{darkgreen}{rgb}{0.35, 0.65, 0.35}
\definecolor{darkred}{HTML}{822522}
\definecolor{darkblue}{HTML}{485DB5}

\hypersetup{hidelinks}

\newcommand{\filledtriangle}{
    \begin{tikzpicture}[scale=.26, line width=0.5pt]
        \fill[black] (0,-0.5) -- (0.866,0) -- (0,0.5) -- cycle; % Triangle definition
    \end{tikzpicture}
}

\newcommand\subfig[1]{(#1)}

\newcommand{\abs}[1]{\left\lvert #1 \right\rvert}
\newcommand{\swap}{\widehat{\text{\textls[-75]{S}\textls[-150]{WAP}}}}

\definecolor{RoyalBlue}{rgb}{0.2549,0.4118,0.8824}
\definecolor{ROYALBLUE}{rgb}{0.2549,0.4118,0.8824}

\def\affilA{School of Physics, Trinity College Dublin, Dublin 2, Ireland} 
\def\affilB{Trinity Quantum Alliance, Unit 16, Trinity Technology and Enterprise Centre, Dublin 2, Ireland} 
\def\affilC{Dublin Institute for Advanced Studies, School of Theoretical Physics, Dublin, Ireland} 
\def\affilD{Algorithmiq Ltd, Helsinki, Finland}

\usepackage{cleveref}
\Crefname{equation}{Eq.}{Eqs.}
\Crefname{figure}{Fig.}{Figs.}
\Crefname{section}{Sec.}{Secs.}
\Crefname{supplementary note}{SM-Sec.}{SM-Secs.}
\crefname{paragraph}{Paragraph}{Paragraphs}
\Crefname{paragraph}{Paragraph}{Paragraphs}

\makeatletter
\def\maketitle{
\@author@finish
\title@column\titleblock@produce
\suppressfloats[t]}
\makeatother

\begin{document}

\title{Anomalous transport in U(1)-symmetric quantum circuits}
% maybe an alternative might be: Anomalous Transport in Symmetry-Protected Quantum Circuits

% Last edits before submission:
% - Highlight more finiteness of our analysis
% - Fix last appendix
% - Prethermal

\author{Alessandro Summer}
\email{summera@tcd.ie}
\affiliation{\affilA}
\affiliation{\affilB}

\author{Alexander Nico-Katz}
\affiliation{\affilA}
\affiliation{\affilB}

\author{Shane Dooley}
\affiliation{\affilC}
%\affiliation{\affilB}

\author{John Goold}
\affiliation{\affilA}
\affiliation{\affilB}
\affiliation{\affilD}

\date{Published 17 January 2026; corrected 10 April 2026}

% \begin{abstract}
%  In this work we investigate discrete-time transport in a 
%  generic $\mathrm U(1)$-symmetric disordered model tuned across an array of different dynamical regimes. We develop an aggregate quantity, a circular statistical moment, which is a simple function of the magnetization profile and which elegantly captures transport properties of the system. From this quantity we extract transport exponents, revealing behaviors across the phase diagram consistent with localized, diffusive, and -- most interestingly for a disordered system -- superdiffusive regimes. Investigation of this superdiffusive regime reveals the existence of a prethermal ``swappy'' regime unique to discrete-time systems in which excitations propagate coherently; even in the presence of strong disorder.
% \end{abstract}

\maketitle

\section{Abstract}

In this work, we investigate discrete-time transport in a generic $\mathrm U(1)$-symmetric disordered model tuned across an array of different dynamical regimes. We develop an aggregate quantity, a circular statistical moment, which is a simple function of the magnetization profile and which elegantly captures transport properties of the system. From this quantity, we extract transport exponents, revealing behaviors across the phase diagram consistent with localized, diffusive, and---most interestingly for a disordered system---superdiffusive regimes. Investigation of this superdiffusive regime reveals the existence of a prethermal ``swappy'' regime unique to discrete-time systems in which excitations propagate coherently; even in the presence of strong disorder.

\section{Introduction}
\label{sec:intro}

%%%% PLEASE DON'T DELETE %%%%%%
%\shane{
%Quantum many-body dynamics have traditionally been studied from the perspective of \emph{continuous-time} dynamics in \emph{energy-conserving} models. In practice, digital simulation of these dynamics requires a discrete-time approximation (e.g., via the Suzuki-Trotter expansion~\cite{Tro-59a, Suz-90a}) which breaks the energy conservation and introduces Trotter errors.

%However, driven by the advent of large-scale, high-fidelity digital quantum devices, the \emph{discrete-time} dynamics in \emph{non-energy-conserving} models are increasingly viewed as an interesting topic of investigation in their own right. In other words, Trotter errors are no longer a bug, but are instead regarded as a feature of the dynamics. Interesting many-body phenomena that arise in the this class of models include: discrete time crystals~\cite{Khe-16a, Mi-21a}, 
% driven by digital Floquet kicks and stabilized by disorder, 
%measurement induced phase transitions~\cite{Ski-19a, Noe-22a, Goo-23a},
% also with Clifford circuits~\cite{Nir-24a},
% many body scars~\cite{Log-24a}, 
% Hilbert space fragmentation, 
%and cellular automata~\cite{Pir-20a}.

%In energy conserving models with an additional $\mathrm U(1)$-symmetry, one of the most useful ways of characterizing the dynamics is through transport properties... By breaking the energy-conservation but preserving the $\mathrm U(1)$-symmetry we can ask if any novel features emerge.
%}
%%%% PLEASE DON'T DELETE %%%%%%

The advent of large-scale, high-fidelity, tunable, and accessible digital quantum devices is revolutionizing modern quantum physics. Notable advances include the implementation of large-scale quantum algorithms, the simulation (via e.g., the Suzuki-Trotter expansion~\cite{Tro-59a, Suz-90a}) of continuous-time models on actual hardware~\cite{Ber-15a, Low-17a, Cam-19a, Bar-21a, Fau-24a, Ric-21a}, diagonalization algorithms~\cite{Sum-24a, Yos-24a}, and the analysis of transport properties~\cite{Kee-23a, Mi-24a, Gya-24a, Fis-24a}. 

\begin{figure}
\includegraphics[width=.962\columnwidth]{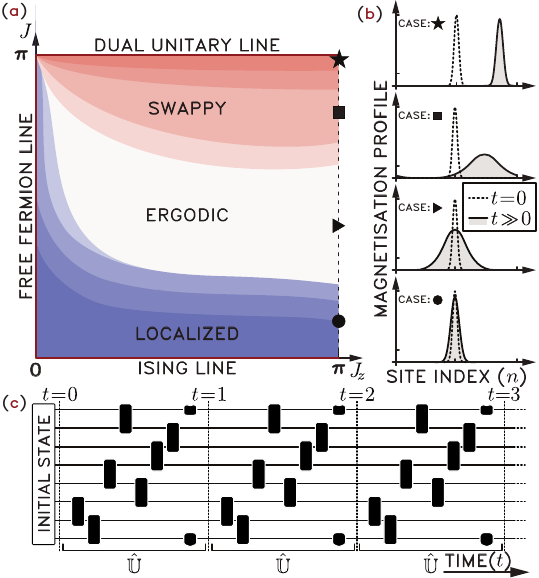}
\caption{\textbf{Schematics of model phases, dynamics, and typical circuit realization.} Schematics showing \subfig{a} different regimes of our generic $\mathrm U(1)$-symmetric model. Along the three solid borders the model is integrable; whilst the dashed border at $J_z=\pi$ spans four distinct regimes, and is the region we predominantly address in this article. \subfig{b} shows the initial (dotted) and late-time (solid) spin magnetization profiles for different phases along the $J_z=\pi$ line. \subfig{c} shows discrete time evolution as determined by a (periodic) $\mathrm U(1)$-symmetric Floquet unitary $\hat{\mathbb{U}}$ comprised of nearest-neighbor gates.} 
\label{fig:schematic}
\end{figure}

However, these advances have also resulted in an emerging reconceptualization of large circuits as realizing digital phases of matter \textit{in their own right}. Notable examples include: discrete time crystals~\cite{Khe-16a, Mi-21a}, 
% driven by digital Floquet kicks and stabilized by disorder, 
measurement-induced phase transitions~\cite{Ski-19a, Tur-21a, Noe-22a},
% also with Clifford circuits~\cite{Nir-24a},
many body scars~\cite{Log-24a, And-24a}, Hilbert space fragmentation~\cite{Mou-22a, Han-24a, Log-24b}, and cellular automata~\cite{Pir-20a}. 
Essentially, despite their fundamental differences, a sufficiently large discrete-time system exhibits emergent phenomena in a manner consistent with continuous phases of matter. This deterritorialization of phase, now encompassing digital systems, yields questions of immediate interest and relevance: how can one probe phase transitions and information-spreading in digital matter? As digital systems do not conserve energy, but can sustain other conserved quantities, what are the transport features of such systems? Are there additional regimes and features unique to digital phases of matter, absent in their continuous-time counterparts?

In this work, we seek to address some of these questions by constructing a generic gate-based, disordered, $\mathrm U(1)$-symmetric model; inspired by the more recent development of a discrete-time version of the XXZ spin chain~\cite{Van-18a, Lju-19a, Lju-19b}. The continuous-time XXZ model, and its isotropic counterpart, the Heisenberg model, were conceived as purely theoretical constructs~\cite{Bet-31}, but have become standard tools for investigating transport properties~\cite{Ber-16a, Zni-16a, Pro-11a} and many-body localization (MBL)~\cite{Pal-10a, Lui-15a, Aba-19a, Nic-22b, Nic-22a, Ser-14a, Ser-15a, Khe-17a, Lui-20a, Nic-23a, Col-24a}. Meanwhile, the transport properties of its discrete-time counterpart remain only partially mapped out. 

Our model can be tuned across a two-dimensional phase space to exhibit both integrable and non-integrable regimes, and a range of emergent phenomena. This includes a crossover between ergodic and localized regimes~\cite{Pro-98a, Pro-99a, Pro-02a, Sie-23a}, and a line along which all gates in our model are dual-unitary (DU)~\cite{Gop-19a, Ber-19a}. These DU gates are endowed with an additional $\mathrm U(1)$ symmetry, and thus become generalized SWAP gates (see \Cref{sec:overview}). 

We leverage our model into an investigation of the static properties of our model, revealing no unexpected behavior. However, we subsequently interrogate transport in our model and identify a prethermal regime that is \textit{a priori} unique to discrete-time systems.
In the vicinity of the DU line, excitations propagate faster than in the ergodic phase, leading to transient, coherent information transfer over short timescales, the ``swappy'' regime. 
This regime is intrinsically Floquet in origin: it arises from the finite, non-scaled unitary kicks that define the discrete-time dynamics, and therefore disappears in the continuous-time (Trotterized) limit where the kick amplitude is taken to zero. In this sense, the swappy regime has no analog in continuous-time XXZ models and reflects a genuinely discrete-time transport phenomenon.
Finally, by deploying circular moments related to wrapped probability distributions, we develop an aggregate quantity, $R$, which captures the features of many-body transport in a concise way. As this quantity is singularly composed of local $Z$-basis expectation values, it is highly amenable to experimental implementation; paving the way for the near-future experimental analysis of transport in digital matter.\\

In \Cref{sec:model}, we present our model and compare it with similar ones. \Cref{sec:overview} outlines the primary features anticipated to appear in our phase diagram, as suggested by the existing literature. \Cref{sec:results} reports the results: \Cref{sec:spectral_properties} for static properties obtained via exact diagonalization and \Cref{sec:transport_properties} for dynamical properties that define the finite time and size transport analysis of our model.

%{\color{red} I think the introduction needs a bit of work - i would not motivate at all from quantum computing but rather from the inspiration of quantum simulation field into condensed matter towards discrete time systems with no trotter error. This has lead to new digital phases of matter such as DU circuits and measurement based transition etc. Then i would say another way to approach the problem is discretising canonical hamiltonian problems such as XXZ ising etc and asking what is the essential difference in physics and then you say a topic of interest is MBL cite all circuit papers and indeed the role of symmetries. Then you motivate what we do. i.e. study the role of U(1) symmetry in disordered circuit and the emergent high temperature transport. Then summarize a little the findings. I don't think its so much work.  }

\section{Results}
\subsection{Model}
\label{sec:model}
%MODEL(A): technical aspects of the model
We consider a system of $N$ qubits undergoing discrete-time evolution under the repeated application of a Floquet unitary $\hat{\mathbb{U}}$. The Floquet unitary is decomposed as a product of $N$ nearest-neighbor gates $\hat{\mathbb{U}} = \prod_{\{P(n)\}} \hat{U}_{n,n+1}$, ordered according to a random permutation $P$ of the set of all site indices $n$ (illustrated in \Cref{fig:schematic}\subfig{c}). We assume periodic boundary conditions. 
The two-qubit gates have the form: 
\begin{equation} 
\hat{U}_{n,n+1} = e^{-i\hat{H}_{n,n+1}} e^{-i h_n \hat{S}_n^z } e^{-i h'_n \hat{S}_{n+1}^z } , 
\label{eq:two-qubit-gate} 
\end{equation}
which is composed of the single-qubit rotations $e^{-i h_n \hat{S}_n^z } e^{-i h'_n \hat{S}_{n+1}^z }$ and a nearest-neighbor XXZ-like interaction generated by the Hamiltonian: 
\begin{equation}
\label{eq:ham-xxz} 
\hat{H}_{n,n+1} = \frac{J}{2} \left( \hat{S}_n^+ \hat{S}_{n+1}^- e^{i\phi_n} + \hat{S}_n^- \hat{S}_{n+1}^+ e^{-i\phi_n} \right) + J_z \hat{S}_n^z \hat{S}_{n+1}^z , 
\end{equation}
with the Peierls phase $\phi_n$~\cite{Pei-93a}. Here, the operators $\hat{S}_n^\alpha$ for $\alpha \in \{x,y,z\}$ are the standard spin-$1/2$ operators acting locally on the $n$-th site, and the raising and lowering operators are defined as $\hat{S}^\pm_n = \hat{S}^x_n \pm i \hat{S}_n^y$ respectively. The two-qubit gate in \Cref{eq:two-qubit-gate}, with five parameters $\{h_n, h'_n, \phi_n, J, J_z \}$, can, in fact, describe any two-qubit gate that conserves the total magnetization $\hat{M} = \sum_{n}\hat{S}_n^z$ in the $z$-direction 
%\shane{[REF]} \aleS{Not sure what kind of ref needed here}
. Since each individual gate conserves the magnetization, the global Floquet unitary also conserves the total magnetization $[\hat{\mathbb{U}}, \hat{M}] = 0$, which is an essential prerequisite to define the transport of spin excitations.

For each two-qubit gate $\hat{U}_{n,n+1}$ we sample the three phase parameters $\{ h_n, h'_n, \phi_n \}$ randomly from the uniform distribution on the interval $[-\pi, \pi]$. This introduces disorder to our circuit model. The two remaining parameters $J$ and $J_z$, however, are fixed across all gates in a given Floquet unitary. Tuning $J$ and $J_z$ allows us to explore various regimes of transport in our model. Observing that the two-qubit gate $\hat{U}_{n,n+1}$ is $2\pi$-periodic in both $J$ and $J_z$ we can restrict our parameter space through patterning (reflection and tessellation) to the region $J, J_z \in [0, \pi]$, shown in \Cref{fig:schematic}\subfig{a}.

Similar discrete-time models have appeared in recent literature~\cite{Lev-15a, Sun-18a, Mor-22a, Hah-23a, Sht-23a, Lon-23a, Khe-18a, Rak-18a, Sie-23a, Jon-24a}. Notably, Refs.~\cite{Khe-18a, Rak-18a, Sie-23a, Jon-24a} use $\mathrm U(1)$ symmetric models but with a brickwork geometry. Our random gate ordering is more similar to the approach in Ref.~\cite{Mor-22a}. This approach also allows systems with odd and even site numbers, $N$, to be treated equivalently and avoids issues related to the slower thermalization of brickwork circuits, as discussed in Ref.~\cite{Jon-24a}. A more in-depth analysis of gate permutations is presented in \Cref{supp:typical_drift} of the Supplementary Materials (SM). Finally, we note that our model is well-suited to for realization on currently available quantum hardware. A different gate decomposition of $\hat{U}_{n,n+1}$, which may be more suitable for implementation on quantum simulators, is discussed in \Cref{supp:pauli_to_peierls}.

\subsection{Overview of regimes of transport dynamics}
\label{sec:overview}
Before presenting our numerical results, in this section, we give an overview of the expected different regimes of dynamics in our model. First, we identify three solvable regimes of our model, corresponding to three solid borders of the $J, J_z \in [0, \pi]$ square in \Cref{fig:schematic}\subfig{a}:
\begin{enumerate} 
\item If $J_z = 0$ the Floquet unitary is integrable, since it can be mapped to a quadratic fermion model by a Jordan-Wigner transformation. 
\item If $J=0$ the Floquet unitary reduces to an Ising model in a disordered longitudinal magnetic field, since it is completely diagonal in the $\hat{S}_n^z$ basis of each spin. 
\item If $J = \pi$ each gate of our model is DU. In what follows we show that imposing $\mathrm U(1)$ symmetry to DU gates leads to generalized SWAP gates making the system effectively non-interacting and therefore integrable.
\end{enumerate}

In the first of these regimes, along the free-fermion line $J_z = 0$, we expect transport to be suppressed by Anderson localization, as a result of the disorder fields $\{ h_n, h'_n, \phi_n \}$. Similarly, in the second of the exactly solvable regimes, along the Ising line, there is no transport of spin-$z$ excitations since, at $J=0$, not only is the total magnetization $\hat{M}$ conserved, but also the local magnetization $\hat{S}_n^z$ for every spin $n$ is conserved.

We now turn our attention to the third of our exactly solvable regimes, at $J = \pi$, where the two-qubit gate can be decomposed as:
\begin{equation}
\begin{split}
\hat{U}_{n,n+1} =& e^{-i\kappa_{+}} \ket{\!\uparrow\uparrow} \bra{\uparrow\uparrow\!} + e^{-i\xi_{+}} \ket{\!\downarrow\uparrow} \bra{\uparrow\downarrow\!} \\ 
& + e^{-i\xi_{-}} \ket{\!\uparrow\downarrow} \bra{\downarrow\uparrow\!} + e^{-i\kappa_{-}} \ket{\!\downarrow\downarrow}\bra{\downarrow\downarrow\!} ,
\label{eq:U_swap}
\end{split}
\end{equation}
for $\kappa_{\pm} = \mp(h+h')/2 +J_z/4$ and $\xi_{\pm} = \pi \pm(h - h')/2 - J_z/4 \mp\phi$. We see that, along this line in parameter space, our two-qubit gate becomes a SWAP gate that also imprints a phase on the swapped particles. We will refer to this gate as a generalized SWAP and to the $(J,J_z)=(\pi,\pi)$ point as the ``SWAP point''. This use of the term is justified within the context of the Weyl chamber, where gates connected by local rotations are defined as equivalent~\cite{Zha-03a, Hah-23a}. 
% The Weyl chamber is a geometric tool used to represent and classify two-qubit gates by mapping them to a tetrahedral region, where each gate is characterized by three coordinates (couplings along different directions).
% \shane{ToDo: comment that along this $J=\pi$ line the gate is dual-unitary... Recently, dual-unitarity has attracted a high degree of interest, primarily because it was shown that circuits with dual-unitary gates admit exact solutions to some quantities that would ordinarily be very difficult to calculate. Interestingly, such exact calculations are even possible when the dual-unitary circuit is chaotic. Add more details + references about dual-unitarity.}
Hence, we do not expect the system to thermalize along the DU line, as any spin excitation propagates through the circuit unimpeded via a series of these generalized SWAP gates. The intrinsic behavior of this type of gate ensures that any DU circuit preserving $\mathrm U(1)$ symmetry is integrable, regardless of connectivity.\\

The non-thermalizing behavior along these three borders raises several interesting questions about the dynamics if we vary slightly away from them.
Traditionally, this has been studied in the proximity of the Ising line. Even if we perturb away from it, by increasing $J$ to small but non-zero values, spin transport is still suppressed, since the kinetic energy $J$-term in \Cref{eq:ham-xxz} is dominated by the disorder fields $\{h_n, h'_n, \phi_n \}$. This is roughly analogous to MBL in continuous-time models~\cite{Aba-19a, Sie-24a}, but, as MBL is not the main focus of this article, we do not interrogate the stability of this regime in the thermodynamic limit. 
However, one can wonder if something similar can arise around the DU line. Will the system thermalize near it? Is there an extended regime of anomalous transport dynamics near the DU line? These are some of the key motivating questions of our work. We will show below that indeed there is a distinctive regime of anomalous transport proximate to the DU line, which we will call the ``swappy'' regime.
Moving further away from the three solvable sides of our $J, J_z \in [0, \pi]$ square, we enter the ergodic regime. Here, we expect transport of spin excitations to lead to rapid thermalization of the system. %\shane{Previous studies on continuous-time models have found... diffusive transport of spin excitations... In the ergodic regime, does our discrete-time model also exhibit thermalization through the diffusion of spin excitations?} 

To explore the behavior of our model between the Ising and the DU line, we focus on the $J_z = \pi$ line, since it is maximally distant (in $(J,J_z)$ parameter space) from the integrable free-fermion line, and exhibits a wide array of interesting behaviors as a function of the single parameter $J$. In particular, throughout this work we focus on four points representative of the different regimes along this line: localized at $J = 0.395$ (represented by the symbol \ding{108}), ergodic at $J = 1.374$ (represented by \filledtriangle), swappy at $J = 2.551$ (represented by $\blacksquare$), and near-SWAP at $J = 3.138$ (represented by $\bigstar$). These regimes are shown schematically in \Cref{fig:schematic}\subfig{b}. These regimes exhibit different kinds of emergent transport and, as we flexibly refer to regimes and types of transport throughout this work, it is instructive to preemptively summarize our findings here. We find either complete localization or sub-diffusive behavior in the localized regime, diffusive behavior in the ergodic regime, and super-diffusive behavior in the swappy and near-SWAP regime. This super-diffusive behavior is augmented by ballistic propagation of individual excitations in the swappy and near-SWAP regimes. These results are shown in \Cref{fig:variance_drift} and \Cref{fig:finite_size_scaling_analysis}, and discussed in \Cref{sec:extracting_transport_properties} and \Cref{sec:scaling-analysis} respectively. 

We emphasize that a full scaling analysis of the localized regime falls outside the scope of this work, and due to analogous literature surrounding the stability of continuous-time MBL, we expect this question to be intractable given current numerical and experimental capabilities~\cite{Pan-19a, Sie-24a}. We can thus only claim to find phenomenological signatures of localization in this work: i.e., a total breakdown of transport, or the sub-diffusive (and/or logarithmic) spreading of information. These phenomenological signatures may be prethermal effects, or finite-size effects, or both; we defer a more detailed investigation of the stability of localization in discrete-time systems to future study. In this work, we focus on phenomenological signatures of transport throughout the phase diagram, at intermediate system sizes $N \leq 22$, and exponential timescales; with an additional focus on interrogating the nature and stability of the swappy regime. 

\subsection{Analysis of the results}
\label{sec:results}

Our goal in this paper is to understand the various equilibrium and dynamical regimes in our $\mathrm U(1)$-symmetric circuit model, outlined in the previous section. To characterize our model, we take two complementary approaches: we compute (\Cref{sec:spectral_properties}) static spectral properties and (\Cref{sec:transport_properties}) dynamical properties. These two approaches are complementary, since they reveal different aspects of the model, as described below.

\subsubsection{Spectral properties}
\label{sec:spectral_properties}

\begin{figure}
\includegraphics[width=\columnwidth]{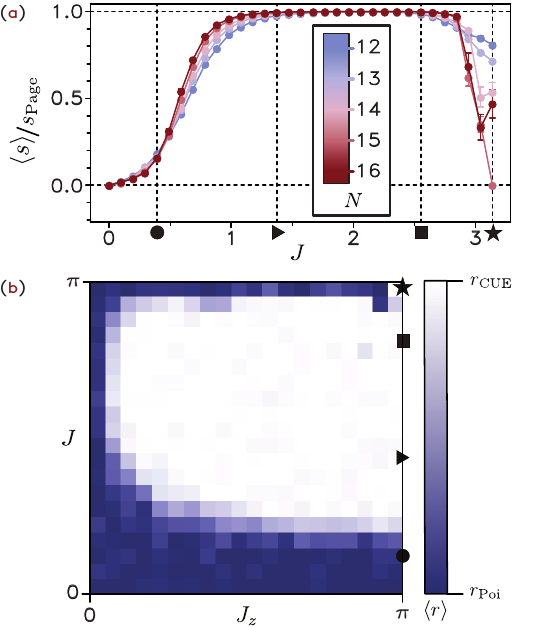}%
\caption{\textbf{Static properties of the model.} \subfig{a} shows results for the averaged von Neumann entanglement entropy normalized by the Page value of Ref.~\cite{Bianchi2019}, along the $J_z = \pi$ line and for various system sizes. Error bars represent standard deviations over at least $40$ disorder realizations, selected with a $3.5\times$ Interquartile Range~\cite{Dekking2005}.
Residual large fluctuations near the SWAP point arise from near-degeneracies in the Floquet spectrum. Panel~\subfig{b} displays the phase diagram for the gap ratio obtained for $N = 15$ with values averaged over $10$--$20$ trajectories.
Results obtained with the POLFED eigensolver in fixed-$M\!=\!0$ sector~\cite{Sie-20b} (see \Cref{supp:polfed} for numerical details).
}
\label{fig:static-properties}
\end{figure}

One way to characterize our model is by means of its spectral properties, i.e., properties relating to the eigenphases and eigenvectors of the Floquet unitary, $\hat{\mathbb{U}} \ket{\varphi_i} = e^{-i\varphi_i} \ket{\varphi_i}$. Specifically, we study the mean eigenstate entanglement entropy and the mean eigenvalue gap ratio--both indicators of quantum chaos~\cite{Sie-24a}. 
Both these quantities require partial diagonalization of $\hat{\mathbb{U}}$, which we perform using the POLFED method~\cite{Sie-20b}, technical details are given in \Cref{supp:polfed} of the Supplementary Materials (SM).
We briefly discuss the relevance of these quantities to the analysis of the ergodicity of a system in \Cref{supp:infinite_time_diagnostics} of the SM.

First, consider a bipartition of the system into subsystems $A$ and $B$. The entanglement entropy of a Floquet eigenstate $\ket{\varphi_i}$ is given by:
\begin{equation}
    s_i = -\Tr[\hat{\rho}_i^A \ln (\hat{\rho}_i^A)] ,
    \label{eq:entanglement_entropy}
\end{equation}
where $\hat{\rho}_i^A = \Tr_B[\ket{\varphi_i} \bra{\varphi_i}]$ is the reduced density matrix for subsystem $A$. Averaging $s_i$ over eigenstates in a fixed magnetization sector $M$ (see \Cref{supp:methods} of the SM) and across different Floquet unitaries yields the mean entanglement entropy $\langle s \rangle$. For an ergodic system, we expect the eigenstates to have an average entanglement entropy similar to a typical random state in the magnetization sector, i.e., close to the Page entropy~\cite{Pag-93a}.
%$s_\text{Page} = \ln d_A - d_A^2/2d$, where $d$ is the total Hilbert space dimension and $d_A$ is the dimension of subsystem $A$ \cite{Pag-93a}, as discussed in \Cref{supp:methods}. %here, $d = d^N_M$ and $d_A = d^{N/2}_M$. 
Specifically, as our numerics are performed in a fixed total magnetization sector $M=0$, the natural random-state benchmark is the Page value restricted to that charge sector. 
The exact finite-size expression has been derived by Bianchi and Donà in Ref.~\cite{Bianchi2019},
\begin{equation}
s_\text{Page}=\sum_{N_A}\bar\lambda_{N_A}\bar s_{N_A}-\sum_{N_A}\bar\lambda_{N_A}\big[\psi(d_{N_A}+1)-\psi(D+1)\big],
\label{eq:right_page_value}
\end{equation}
with $d_{N_A}=\dim\mathcal H_A^{(N_A)}$ ($d_{N_A}=\binom{L_A}{N_A}$), $d_{N_B}=\dim\mathcal H_B^{(M-N_A)}$, $D=\sum_{A}d_{N_A}d_{N_B}$ and $\bar\lambda_{N_A}=d_{N_A}d_{N_B}/D$. 
We therefore normalize the mean eigenstate entropies by this fixed-magnetization Page value when assessing Page-like saturation in the ergodic regime.
For this reason, it is convenient to normalize the average entropy by the Page entropy, $\langle s \rangle / s_\text{Page}$.
In \Cref{supp:entropy_computation} of the SM, we discuss how to compute the entanglement entropy of the eigenstates efficiently.
In \Cref{fig:static-properties}\subfig{a} we see that, for our model, the mean entanglement entropy varies from a small value in the localized regime ($J \ll \pi$) to the Page entropy in the ergodic regime. This is consistent with the conclusion that -- at least for our finite-$N$ -- the system fails to thermalize in the long-time limit in the localized regime, but thermalizes in the ergodic regime. \Cref{fig:static-properties}\subfig{a} also appears to show deviations from the Page entropy as we approach the SWAP point $J \to \pi$. However, we observe that as the system size $N$ increases, the range of $J$ values around $\pi$ associated with lower entanglement entropy contracts.
We attribute this instability to the emergence of (near-)conserved operators as the Floquet gates approach the DU line. These operators produce near-invariant subspaces and thus effectively split the spectrum into overlapping subspectra. When raw spectral measures are computed without first resolving those (approximate) symmetry sectors, the resulting level-spacing statistics and entanglement values can be distorted~\cite{Ata-13a}.
We comment on this in \Cref{supp:transfer_matrix} of the SM.

Our next quantifier is the mean gap ratio. For each Floquet eigenphase $\varphi_i$, the ratio of consecutive gaps is defined as:
\begin{equation}
r_i = \frac{\min\{ \delta\varphi_{i-1}, \delta\varphi_i \}}{\max\{ \delta\varphi_{i-1}, \delta\varphi_i \}},
\label{eq:gap_ratio}
\end{equation}
where $\delta\varphi_i = \varphi_{i+1} - \varphi_i$ is the gap between consecutive eigenphases within the same magnetization sector~\cite{DAl-14a}. We compute the mean gap ratio $\langle r \rangle$ by averaging $r_i$ over all computed eigenphases in a sector and over many Floquet unitaries. In the absence of time-reversal symmetry, the model is expected to yield $\langle r \rangle \approx r_\text{CUE} = 0.6027$ if the system is ergodic, or $\langle r \rangle \approx r_\text{Poi} = 0.3863$ if the system is integrable~\cite{Ata-13a}. \Cref{fig:static-properties}\subfig{b} shows the transition between the localized regime for small $J$ and the ergodic regime for sufficiently large $J$. It also shows a deviation from the ergodic value when $J_z = 0$, corresponding to our model being integrable (mappable to free fermions), as well as on the dual-unitary line $J = \pi$. 

%Each point represents an average over 20-35 Floquet unitaries. 

Both the mean entanglement entropy and the mean gap ratio probe equilibrium properties of the model in the infinite-time limit. 
These two diagnostics probe complementary aspects of the long-time state. The mean eigenstate entanglement entropy tests whether typical energy eigenstates have thermal reduced states (as expected if the Eigenstate Thermalization Hypothesis, ETH, holds), while the mean gap ratio diagnoses spectral correlations: Wigner-Dyson statistics imply level repulsion and RMT-like eigenvectors, whereas Poisson statistics imply level clustering. Together, therefore, they indicate whether the system's infinite-time (thermal) behavior is consistent with ergodicity or with nonthermal dynamics.

So, even if the system ultimately thermalizes, interesting intermediate-time behaviors may arise that will not be detected by these spectral quantities. Next, we turn to the quantification of the transport properties of the model, which can reveal distinct dynamical regimes at intermediate times.

We remark that the crossing point extracted from the mean adjacent gap-ratio and from the rescaled eigenstate entanglement exhibits a modest, systematic shift as $L$ is increased for the range of sizes accessible to our numerics. Such finite-size drifts are routinely observed in ED studies of disordered interacting chains and were historically used as an operational probe of the putative MBL transition~\cite{Oga-07a}. However, more recent analyses emphasize that these finite-size signatures are difficult to extrapolate to the thermodynamic limit: rare thermal regions and avalanche-type instabilities can produce apparent localization at small sizes, which is destroyed upon increasing scale~\cite{DeRoeck2017,Thiery2018}, and large-scale numerics and reviews discuss slow crossovers and sample-to-sample fluctuations that complicate direct interpretation~\cite{Alet2018,Doggen2021}. Accordingly, while our data for accessible $L$ are consistent with a crossover from ergodic to strongly suppressed transport regimes, they do not constitute proof of a stable MBL phase in the thermodynamic limit; we therefore limit our claims to the finite-system behavior reported here.

\subsubsection{Transport properties}
\label{sec:transport_properties}

We first present the raw numerical data for the evolution of the local magnetization profile, starting from a non-equilibrium initial state consisting of a localized spin excitation in a homogeneous spin background. Suitably quantifying the spread of the initially localized excitation should encapsulate the transport properties of the system. However, we find that the local magnetization profile is inconvenient for extracting transport coefficients. Then, we map the local magnetization profile to a quasi-probability distribution. Using the wrapped normal distribution as a helpful reference model, we use the quasi-probability distribution to extract transport coefficients. We therefore discuss these transport coefficients, and finally, we conduct a scaling analysis of the transport properties. This study reveals an intermediate-time prethermal ``swappy'' regime, unique to digital systems, where information propagates coherently and ballistically, even in a strongly disordered setting. We discuss heuristic mechanisms underlying this regime, and its stability, in \Cref{sec:results}.

\paragraph{Local magnetization profile.}
\label{sec:numerical_magnetization_profile} 

\begin{figure*}
\centering
\includegraphics[width=\textwidth]{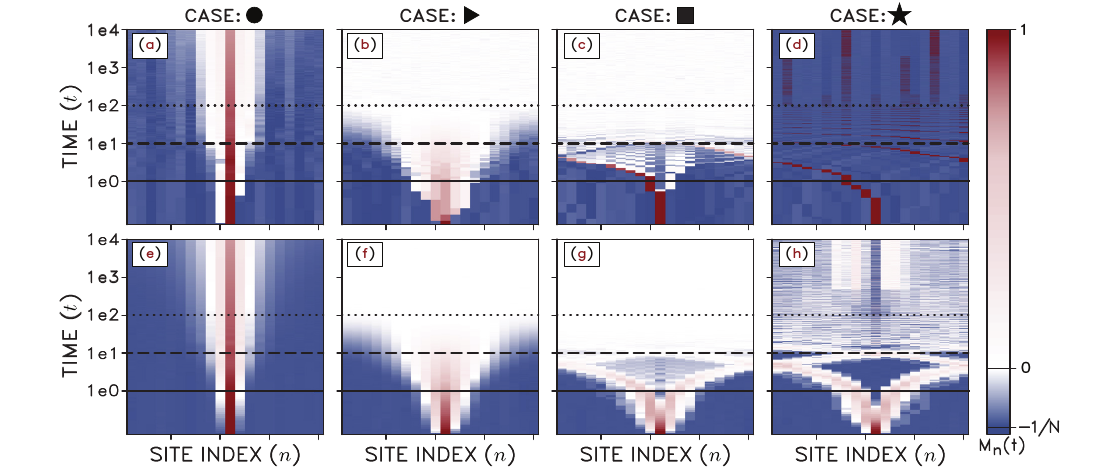}%
\caption{\textbf{Dynamics of a single spin inhomogeneity in the localized, ergodic, swappy, and near-SWAP regimes of the model.} Results of numerical propagation of the initial spin inhomogeneity in the $M=0$ sector under $\mathrm U(1)$-symmetric Floquet dynamics at $N=20$. \subfig{a}--\subfig{d} show typical results for single realizations of the initial state and Floquet unitary within the localized regime, ergodic regime, and two points in the swappy regime, respectively. \subfig{e}--\subfig{h} show results in the respective regimes after averaging over $100$ realizations. Horizontal lines indicate the time taken to complete a single Floquet cycle (solid), $N/2$ (dashed), and the time after which we calculate results stroboscopically (dotted).
Although the light cones in \subfig{b} and \subfig{f} visually show the excitation spreading and an apparent equilibration for $t$ between $\mathcal{O}(N)$ and $\mathcal{O}(N^2)$, we confirm diffusive transport quantitatively in the following sections. An analysis of what happens exactly on the light-cone is offered in \Cref{supp:transfer_matrix} of the SM.}
\label{fig:magnetization_profiles}
\end{figure*}

The central object used to explore dynamics in our model is the local spin magnetization profile
\begin{equation} 
    M_n (t) = \Tr [\hat{\mathbb{U}}^t  \hat{\rho}(0) (\hat{\mathbb{U}}^\dagger)^t  \hat{S}_n^z] . \label{eq:rho_0} 
\end{equation}
We exploit initial states of the form 
\begin{equation} 
    \hat{\rho}(0) =
    \frac{
      \hat{P}_M \left[\ket{\!\uparrow}\bra{\uparrow\!}_{N/2}
      \otimes \left( \frac{\hat{\mathbb{I}}_2}{2} \right)^{\otimes (N-1)} \right] \hat{P}_M
    }{
      \Tr\!\left[
      \hat{P}_M \left[\ket{\!\uparrow}\bra{\uparrow\!}_{N/2}
      \otimes \left( \frac{\hat{\mathbb{I}}_2}{2} \right)^{\otimes (N-1)} \right] \hat{P}_M
      \right]
    } ,
    \label{eq:initial-state}
\end{equation}
where $\hat{P}_M$ is a projector onto the subspace of total magnetization $M = \sum_n M_n$. In numerical practice, to avoid working with the density matrix $\hat{\rho}(0)$ in \Cref{eq:rho_0}, we choose the initial pure state $\hat{\rho}(0) \sim \hat{P}^\uparrow_{N/2} \ket{\psi^{(M)}_{\rm rand}}\bra{\psi^{(M)}_{\rm rand}} \hat{P}^\uparrow_{N/2}$ where $\ket{\psi^{(M)}_{\rm rand}}$ is a random state in the subspace of fixed total magnetization $M$, and $\hat{P}_{N/2}^{\uparrow}$ is the projector onto the excited state of the middle spin. This is numerically more efficient, since we only need to operate on the pure state instead of on the density matrix. In \Cref{supp:typical_state} of the SM, we show in detail that this is equivalent to choosing an initial maximally-mixed state in \Cref{eq:rho_0}, up to some negligible random fluctuations. Notably, one can directly connect the quantity of \Cref{eq:rho_0} computed in a given magnetization subsector with high-temperature correlation functions, as is shown in detail in \Cref{supp:expanded-transport} of the SM.

For this choice of initial state, the central spin is fully polarized such that $M_{N/2}(0) = 1/2$. The excess magnetization $M-1/2$, where $M = \sum_n M_n(t)$, is distributed evenly over all other spins $M_{n \neq N/2}(0) \approx M_{\rm B} = (M - 1/2)/(N-1)$. In other words, the initial local magnetization is in the non-equilibrium configuration: 
\begin{equation} 
    M_n (0) = \frac{1}{2}\delta_{n,N/2} + (1 - \delta_{n,N/2}) M_{\rm B} 
    \label{eq:initial-magnetization-profile} . 
\end{equation} 
As total magnetization $M$ is conserved in our model, and as thermalization would result in the initially localized spin excitation spreading through the system, full thermalization corresponds to the total magnetization $M$ being uniformly distributed across all spins $M_n (t\to\infty) \approx M / N$.  As is standard, we focus on the largest subsector: taking $M = 0$ from this point onward.

We present our numerical results for the magnetization profile $M_n(t)$ for a system of $N=20$ particles in \Cref{fig:magnetization_profiles} at the four key points along the $J_z=\pi$ line (as shown in \Cref{fig:schematic}\subfig{a}). The top panels \subfig{a-d} illustrate the dynamics for a typical individual realization of $\hat{\mathbb{U}}$, while the bottom panels \subfig{e-h} show the dynamics of the local magnetization averaged over many different realizations of the Floquet unitary.

As outlined in \Cref{sec:model}, and depicted in \Cref{fig:schematic}\subfig{c}, time evolution is carried out by repeated application of the Floquet unitary $\hat{\mathbb{U}}$. We evaluate each of the first $t=100$ time steps, but to reach later times we apply clusters of Floquet unitaries together $\hat{\mathbb{U}}^{k}$, which effectively realizes a stroboscopic dynamics with a longer period after $t > 100$. 
This stroboscopic approach leads to an apparent preservation of excitation on some particular sites, however, even at this later times, we expect the excitation to move continuously along the chain.
Throughout \Cref{fig:magnetization_profiles}, horizontal lines mark (solid) $t=1$, the application of the first whole $\hat{\mathbb{U}}$, (dashed) $t=N/2=10$, the time for ballistic excitations to reach the system boundary, and (dotted) $t=(N/2)^2=100$, the time taken for a particle on site $N/2$ to diffuse to the boundary of the system. These three benchmarks help compare different types of transport, where the $t=100$ line serves as a reference for identifying diffusive or anomalous behavior. These benchmarks are heuristic guides for distinguishing ballistic, diffusive and anomalous transport. We make these statements quantitative below via fits of a series of different quantities we extract from the evolution of the magnetization profile of the system. White regions denote homogenized $M_n(t) = 0$, indicating regions in which the state realizes a local identity and has self-thermalized.

In the localized regime (\ding{108}), shown in \subfig{a} and \subfig{e}, the initial spin excitation remains localized for a long-term, preserving its general shape and position in the spin chain. The system exhibits self-averaging at late times (i.e., after the early-time relaxation dynamics), consistent with many-body localization studies~\cite{Lez-19a, Aba-19a, Nan-15a, Ale-18a}. Self-averaging means a system's properties become stable and predictable in large samples without many different realizations. This feature breaks down in phase transitions where disorder causes persistent fluctuations across regions, preventing uniform behavior. The initial excitation spread slowly, likely logarithmically, in line with what is usually marked as localized behavior~\cite{Sie-24a, Sie-22b, Pan-19a, Nan-15a}.

In the ergodic regime (\filledtriangle), shown in \subfig{b} and \subfig{f}, diffusive transport is qualitatively evidenced by the excitation reaching the boundary, and the subsequent onset of thermalization, at time $t=100$ (the dotted black line). Also at this point, self-averaging occurs, with individual realizations \subfig{b} resembling the averaged behavior \subfig{f}.

The swappy regime ($\blacksquare$) is displayed in panels \subfig{c} and \subfig{g}. The individual realization in panel \subfig{c} demonstrates a clear leftward propagation of the initial excitation, with the direction determined randomly by the permutation $P$ of the local gates making up the Floquet unitary. Initially, before the small deviations from the SWAP gate become evident, the excitation propagates coherently and ballistically through the system, moving a fixed number of sites per Floquet cycle. We will later quantify this behavior using the instantaneous speed parameter $\overline{\langle\nu\rangle}$ in \Cref{sec:scaling-analysis}. These ballistic excitations eventually decohere, leading to thermalization after roughly $10$ Floquet applications. This fast thermalization is more apparent in the averaged dynamics of panel \subfig{g}, where excitations propagate in both directions, splitting into ``twin peaks.'' These excitations self-interact and thermalize rapidly after $t \sim 10$. This finding is significant: thermalization occurs on a timescale consistent with \textit{ballistic} transport in a disordered quantum system. Such anomalous transport has been observed in disorder-less models (see Refs.~\cite{Van-18a, Lju-19a, Lju-19b}), but here we find counter-intuitive evidence of such behavior in strongly disordered systems.

Near the SWAP point ($\bigstar$), panel \subfig{d} clearly shows the leftward propagation of the initial excitation. However, in this case, the system continues to exhibit a near-perfect exchange of spin excitations with neighboring sites, even at late times. Panel \subfig{h} appears to show thermalization at late times. However, this is actually the result of averaging out the peak, caused by small variations in the underlying propagation speeds. Essentially, while the center of $M_n(t)$ shifts, its initial delta-peaked functional form remains unchanged. Averaging many such randomly centered delta-peaked distributions creates the appearance of thermalization at late times. Once again, we observe an early-time ``twin peaks'' structure, where the excitation propagates coherently and ballistically in both directions. Unlike in most of the swappy regime, the twin peaks here are well-resolved. The absence of full thermalization at the near-SWAP point is likely due to the timescales accessed; with a complete failure to thermalize at all times associated only with the SWAP point $J=\pi$. %\aleS{The peculiarity of this point is therefore that individual trajectories exhibit slow thermalization, while averaging over multiple realizations results in a system that effectively thermalizes at a much faster rate.}

% cite \cite{Sch-20a, Aha-96a, Wis-95a} potentially as referring to typical self-averaging in localized/ergodic systems

% As a precursor to the construction of an appropriate quantifications of spreading, we

%The background magnetization profile $M_{n \neq N/2}(0) \approx M_{\rm B} = (M - 1/2)/(N-1)$ is negative whenever $M \leq 0$

\paragraph{Mapping local magnetization to a quasi-probability.}
\label{sec:quasi-prob}

\begin{figure}
    \centering
    \includegraphics[width=\linewidth]{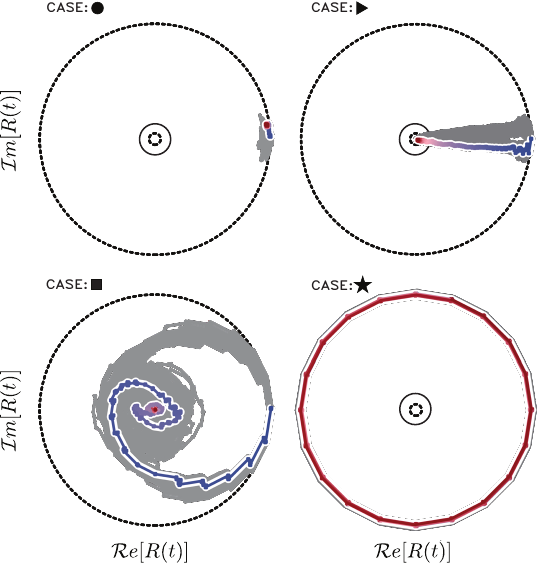}
\caption{\textbf{Dynamics of the circular mean $R(t)$ in the localized, ergodic, swappy, and near-SWAP regimes.} Representations of $R(t)$ on the complex plane, within the unitary circle, are shown for the following values of $J$: (\ding{108}) $J = 0.395$ (localized), (\filledtriangle) $J = 1.374$ (ergodic), ($\blacksquare$) $J = 2.551$ (swappy), and ($\bigstar$) $J = 3.138$ (near-SWAP regime). A typical trajectory from the total $\Omega = 100$ trajectories is highlighted with a gradient from blue to red, corresponding to $t$ from $0$ to $100$, and $N = 20$. The remaining trajectories are displayed in gray.}
    \label{fig:spirals}
\end{figure}

To gain a deeper understanding of the transport dynamics of the initially localized excitation: we first construct a statistical moment derived from the magnetization profile. From this moment, we can extract transport properties (namely transport exponents and drift) which can in turn be related (via an appropriate choice of initial state) to high-temperature correlation functions (see \Cref{supp:expanded-transport} of the SM).

To this end, we first construct a quasi-probability distribution $p_n(t)$ by transforming the magnetization profile $M_n(t)$ as follows: 
\begin{equation}\label{eqn:quasiprob}
    p_n (t) = 2\frac{M_n(t) - M_{\rm B}}{1 - 2M_{\rm B}}.
\end{equation} 
This transformation approximately maps the initial background magnetization values to zero, $p_{n\neq N/2}(0) \approx 0$, with the excitation at $p_{N/2}(t) = 1$, and satisfies Kolmogorov's second probability axiom, $\sum_n p_n(t) = 1$, due to the $\mathrm U(1)$ symmetry (i.e., conservation of $M$). Small fluctuations during the random initial state preparation may cause $M_n (t) < M_{\rm B}$, leading $p_{n\neq N/2}(0)$ to dip below zero. As a result, $p_n(t)$ technically becomes a quasi-probability distribution, violating Kolmogorov's first axiom. However, this effect is minimal and vanishes as $N \to \infty$. We observe no pathological behavior due to these small negative values, so we treat $p_n(t)$ as a valid probability distribution. Full thermalization, $M_n(t\to\infty) = M/N$, results in a uniform distribution, $p_n(t\to\infty) \approx 1/N$.

Since we assume periodic boundary conditions for our model, it is also convenient to introduce the circular mean; defined by the complex number:
\begin{equation}
    R(t) = \sum_n p_n(t) e^{i \theta_n} = e^{i\mu(t)-\sigma(t)^2/2} , \label{eq:R_vec}
\end{equation}
where $\theta_n = \frac{2\pi}{N}\left(n-\frac{N}{2}\right) \in [-\pi, \pi)$ maps the magnetization profile, via the quasi-probability distribution $p_n(t)$, of $N$ spins to a single complex value within the unit circle $|R(t)|\leq 1$. Here, the parameters: 
\begin{equation}\label{eq:extract-params}
    \mu(t) = \text{Arg}[R(t)], \qquad  \sigma(t) = \sqrt{-2\text{log}[|R(t)|]} \geq 1 ,
\end{equation}
quantify the position and the spread of the spin excitation, respectively. To see this intuitively, consider the example of a quasi-probability profile $p_n = \delta_{n,m}$ for a spin excitation localized at site $n=m$. It corresponds to the point $R = e^{i\theta_m}$ on the unit circle, with the argument $\mu = \theta_m$ reflecting the position $m$ of the excitation and spread parameter $\sigma = 0$ indicating that the excitation is perfectly localized. At the opposite extreme, for the example of a completely delocalized spin excitation we have the quasi-probability profile $p_n = 1/N$. It corresponds to the point $R = 0$ at the origin of the unit circle, with the spread parameter $\sigma = \infty$ indicating that the excitation is completely delocalized (the argument $\mu$ is undefined in this case, since the excitation spread uniformly across the chain has no well-defined position). We can thus interpret $R(t)$ as a vector on the complex plane, where drift of the localized excitation corresponds to rotation of $R(t)$ about the origin, and delocalization corresponds to a shrinking of the magnitude of $R(t)$. Taken together, these allow us to conveniently visualize different kinds of transport dynamics on the unit circle: diffusion and thermalization cause $R(t)$ to move towards the origin, and coherent transport of an excitation through the system is realized as rapid rotation of $R(t)$ around the origin. 

Our decision to characterize transport via the circular mean $R(t)$, and via $\mu(t)$ and $\sigma(t)$ as defined in \Cref{eq:extract-params} (rather than linear first and second moments of $p_n$~\cite{Ste-09a, Lev-15a, Ber-21a}) is based on exploiting wrapped normal distributions as possible descriptions of systems with periodic boundary conditions. Our reasoning is as follows: the spreading of excitations in many-body systems is commonly modeled by a Gaussian distribution~\cite{Ber-77a, Roy-18a}. This finds good correspondence with the classical diffusion equation, which in the one dimensional system is 
\begin{equation}
\label{eq:diffusion_equation}
    \frac{\partial f}{\partial t} = D \frac{\partial^2 f}{\partial x^2}.
\end{equation}
Here, an initial delta-function $f=\delta(x-x_0)$ (analogous to our initial state of \Cref{eq:initial-state}) experiences a decay of its peak $\sim t^{-1/2}$ and growth of the standard deviation as $\sim t^{1/2}$. Our model has periodic boundary conditions such that when the excitation, and thus the tails of the associated quasi-probability distribution, reach the boundary, they ``wrap around'' the edge and self-interact. The dynamics of the quasi-probability distribution can thus heuristically be captured by a wrapped normal distribution of the form:
\begin{equation}
    \mathcal{N}_\text{W}(n; \mu, \sigma) \propto \sum_{k=-\infty}^{\infty} \exp{\left[ \frac{-(\theta_n - \mu + 2\pi k)^2}{2\sigma^2}\right]} ,
    \label{eq:wrapped_normal}
\end{equation}
where the summation over $k$ wraps the underlying normal distribution around the boundary, capturing the interactions of particles that diffuse across boundaries. For such a distribution, the bare parameters $\mu$ and $\sigma$ in \Cref{eq:wrapped_normal} are extracted precisely from the circular mean of \Cref{eq:R_vec} via \Cref{eq:extract-params}. A potential limitation of this approach is that the connection between $\sigma(t)$ and an underlying spread parameter of \Cref{eq:wrapped_normal} only holds for systems that realize an approximately wrapped Gaussian form for their magnetization profiles. From \Cref{fig:magnetization_profiles} we see that this connection holds everywhere except in \subfig{c} and \subfig{g}: the swappy regime, which exhibits the staggered patterning at early times (see \Cref{sec:prethermal}). In this case, we instead exploit the mean position $\mu(t)$, from which we extract a drift speed, to characterize the swappy regime.

In \Cref{fig:spirals} we show the dynamics of $R(t)$ for our the four regimes of our model for the first $t=100$ time steps. We see that for the localized regime (\ding{108}), there is no transport of any kind, $R(t)$ remains close to its initial value for all times and trajectories. In the ergodic regime (\filledtriangle), the trajectory exhibits rapid thermalization, $|R(t)| \to 0$, but no coherent information transport occurs. In the swappy regime ($\blacksquare$), coherent information transport is observed, as $\text{Arg}[R(t)]$ varies at a relatively constant rate, in tandem with eventual thermalization, as $|R(t)| \to 0$. Finally, in the near-SWAP regime ($\bigstar$), coherent information transport persists at a constant rate even at late times, with no noticeable equilibration, and $|R(t)|$ remains equal to 1 throughout.

% https://ocw.mit.edu/courses/18-354j-nonlinear-dynamics-ii-continuum-systems-spring-2015/457b849472ce32022b3c69c60fe125ad_MIT18_354JS15_Ch6.pdf

\paragraph{Extracting transport properties. }
\label{sec:extracting_transport_properties}

We are now prepared to quantitatively characterize the transport properties across our four distinct dynamical regimes. To assess the drift of the initial excitation, which is notably pronounced in the swappy and near-SWAP regimes, we define an instantaneous drift speed: 
\begin{equation} 
\label{eq:extract-speed-main-text}
\nu(t) = \left|\frac{d\mu(t)}{dt}\right| 
\end{equation} 
which, in practice, we modify slightly to account for discontinuities when excitations cross the boundary (see \footnote{In practice, we find it more effective to evaluate coherent transport via an instantaneous drift \textit{speed}, $\nu(t)$, derived from a modified mean parameter, $\tilde{\mu}(t)$, as follows:
\begin{equation}\label{eq:extract-speed}
    \nu(t) = \left|\frac{d\tilde{\mu}(t)}{dt}\right|,\qquad
    \tilde{\mu}(t) = \text{Arg}\left[\sum_n p_n(t) e^{i|\theta_n|}\right],
\end{equation}
where this modification restricts $\abs{\theta_n} \in [0,\pi]$ and addresses discontinuities in $\mu(t)$ that arise as excitations cross the boundary's coordinate discontinuity. Numerical evidence supporting the presence of these discontinuities is presented in \Cref{fig:spirals}.}).
To further quantify the spread of the excitation, most prominently observed in the ergodic regime, we fit the parameter $\sigma(t)$ to the form:
\begin{equation} 
\sigma(t) \sim t^{\alpha_\sigma} , 
\end{equation}
and determine the power-law exponent ${\alpha_\sigma}$.

To quantify the decay of the initial excitation due to its spread through the system, we track the maximum value $p_\text{max}(t) = \max_n p_n(t)$. This is fit to the form 
\begin{equation}
p_\text{max}(t) \sim t^{-\alpha_p},
\end{equation} 
where $\alpha_p$ indicates the decay rate. As discussed in \Cref{sec:numerical_magnetization_profile}, conventional diffusion corresponds to $\alpha_\sigma$ (or $\alpha_p$) $= 1/2$, while ballistic transport aligns with $\alpha_\sigma$ (or $\alpha_p$) $= 1$. Anomalous diffusion occurs with $\alpha_\sigma$ (or $\alpha_p$) $< 1/2$ for subdiffusion and $> 1/2$ for superdiffusion.

%The quantity $p_\text{max}(t)$ thus provides a coarse-grained measure of system spreading, similar to $\sigma(t)$.

\begin{figure*}
\centering
\includegraphics[width=\textwidth]{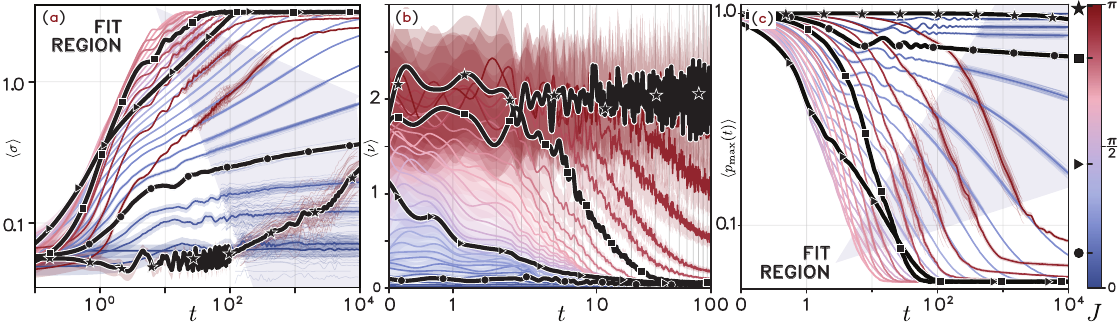}
    \caption{\textbf{Realization-averaged dynamics of transport quantities along the $J_z=\pi$ line.} The dynamics of the realization-averaged parameters for various values of $J \in (0, \pi)$, with $N = 20$ over 100 trajectories, is illustrated for: \subfig{a} the spread $\langle \sigma(t) \rangle$, \subfig{b} drift $\langle \nu(t) \rangle$, and \subfig{c} decay of $\langle p_\text{max}(t) \rangle$. Four distinct regimes are marked with bold lines and symbols: (\ding{108}) localized, (\filledtriangle) ergodic, ($\blacksquare$) swappy, and ($\bigstar$) near-SWAP. The gray-shaded areas in \subfig{a} and \subfig{c} highlight the time intervals used to fit the transport exponents $\alpha_\sigma$ and $\alpha_p$, respectively. Due to stroboscopic sampling limitations, panel \subfig{b} is displayed only up to $t = 100$, effectively capturing all relevant behavior. Error bars are depicted as shaded semi-transparent regions where visible.}
    \label{fig:variance_drift}
\end{figure*}

From each trajectory $R(t)$, we derive the spread $\sigma(t)$ and drift speed $\nu(t)$ as per \Cref{eq:extract-params} and \Cref{eq:extract-speed-main-text}. Averaging over these realizations, denoted by $\langle A\rangle$, yields the sample-averaged spread $\langle\sigma(t)\rangle$ and speed $\langle\nu(t)\rangle$. \Cref{fig:variance_drift} shows these averaged dynamics: \subfig{a} $\langle\sigma(t)\rangle$ and \subfig{b} $\langle\nu(t)\rangle$, for a range of values $J \in (0,\pi)$. The four points of interest are highlighted as bold black lines, labeled by their respective markers. 

% = \frac{1}{\Omega}\sum A
 
In \Cref{fig:variance_drift}\subfig{a}, we compute the spread parameter $\langle \sigma(t) \rangle$ behavior. 
In the localized regime (\ding{108}), information spreads slowly and logarithmically even at late times, analogously to MBL in continuous-time systems. 
In the ergodic regime (\filledtriangle), $\langle\sigma(t)\rangle$ grows as $\sim t^{1/2}$, indicating diffusion until equilibration at $t \approx 10^2$, after which it saturates. The saturation of $\langle\sigma(t)\rangle$ to a constant contrasts with the expected divergence $\langle\sigma(t)\rangle \to \infty$ at equilibrium. However, \Cref{supp:finite-size-plateau} of the SM shows that this saturation is a finite-size effect vanishing as $N \to \infty$.
The swappy regime ($\blacksquare$) exhibits rapid thermalization: initially super-diffusive transport, $\langle\sigma(t)\rangle \sim t$, transitions to diffusion, $\langle\sigma(t)\rangle \sim t^{1/2}$, further explored in \Cref{fig:variance_drift}\subfig{c}. 
In the near-SWAP regime ($\bigstar$), spread remains minimal, with only a slight late-time increase in $\langle\sigma(t)\rangle$. 

In \Cref{fig:variance_drift}\subfig{b}, the instantaneous transport speed $\langle \nu(t) \rangle$ highlights distinct behaviors across regimes. In the localized regime (\ding{108}), no movement occurs, with $\langle \nu(t) \rangle \approx 0$ throughout. The ergodic regime (\filledtriangle) shows non-zero speeds $\langle \nu(t) \rangle > 0$ at early times $t < 10$, but stabilizes to $\langle \nu(t) \rangle \approx 0$ at later times. The swappy ($\blacksquare$) and near-SWAP ($\bigstar$) regimes both display non-zero speeds for long times, with high values of $\langle \nu(t) \rangle$ persisting to extremely long times as $J \to \pi$ approaches the SWAP point. The peak of $\langle \nu(t) \rangle \approx 2$ aligns with the lightcone speed in brickwork circuits, reflecting the typicality of the permutation $P$ and the Floquet unitary $\hat{\mathbb{U}}$, as discussed in \Cref{supp:typical_drift} of the SM. Calculations are limited to $t < 100$ to avoid unreliable speed estimates at stroboscopically sampled times.\\

In \Cref{fig:variance_drift}\subfig{c}, we analyze the maximum value $\langle p_\text{max}(t) \rangle$. Its behavior resembles that of $\langle \sigma(t) \rangle$ in \Cref{fig:variance_drift}\subfig{a} but inverted: high $\langle p_\text{max}(t) \rangle$ values align with minimal spreading, while low values indicate a lack of spreading of the excitation. We see that, in the localized and near-SWAP regimes, $\langle p_\text{max}(t) \rangle$ remains constant, or decays extremely slowly, since the spread of the excitation is suppressed. In the ergodic and swappy regimes, however, the decay of $\langle p_\text{max}(t) \rangle$ is more rapid, consistent with the rapid spread of the excitation. %Both the slow and rapid decay of $\langle p_\text{max}(t) \rangle$ are seen intuitively in the slow or rapid decay of the absolute value of $R(t)$ to zero in Fig. \ref{fig:spirals}.

We note that, taken together, our three quantities $\langle \sigma (t) \rangle$, $\langle \nu (t) \rangle$ and $\langle p_\text{max}(t) \rangle$ allow us to distinguish the dynamical behaviors in our various regimes. For instance, the spread $\langle \sigma (t) \rangle$ shows very similar behavior for both the swappy ($\blacksquare$) and ergodic (\filledtriangle) regimes, but these two regimes are clearly distinguished by the speed $\langle \nu (t) \rangle$. On the other hand, the speed $\langle \nu (t) \rangle$ does not easily distinguish the dynamical behaviors in the localized (\ding{108}) and ergodic (\filledtriangle) regimes, but these are easily distinguished with the quantity $\langle p_\text{max}(t) \rangle$.

%A notable feature in \Cref{fig:variance_drift}\subfig{c} is its smoothness relative to $\langle \sigma(t) \rangle$, with the crossover observed in the swappy regime (from $\langle \sigma(t) \rangle \sim t^{\alpha_\sigma}$ with $\alpha_\sigma=1$ to $\alpha_\sigma=1/2$) absent in $\langle p_\text{max}(t) \rangle$. \shane{This suggests that the wrapped normal distribution may not effectively characterize early-time dynamics in this regime.}

%\aleS{?} In \Cref{fig:magnetization_profiles}\subfig{c}, just before full thermalization ($1<t<10$), the magnetization profile shows staggering, absent in other regimes. Ignoring the main excitation, the profile appears multi-modal, skewing the circular mean $R(t)$~\cite{Sal-40a, Fis-93a}. This may be due to imperfections in the SWAP gate within the swappy regime, which leave fragments of the excitation during propagation. This behavior resembles a simple Markov model as discussed in \Cref{sec:prethermal}% and warrants further exploration in future studies
.

\paragraph{Extracting exponents.}
\label{sec:scaling-analysis}

\begin{figure}
\centering
\includegraphics[width=\columnwidth]{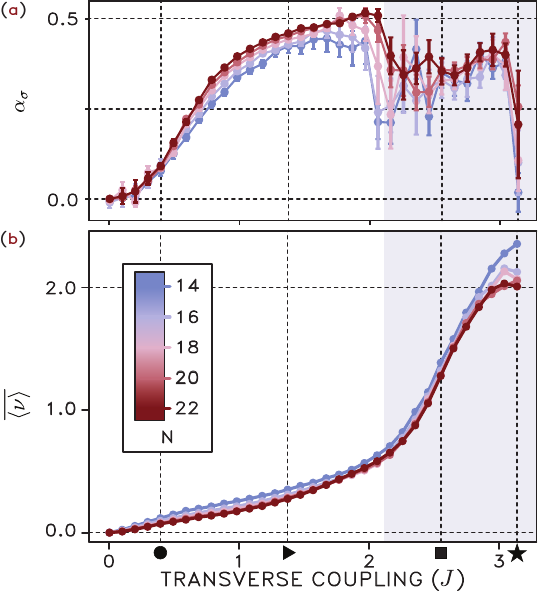}% finite_size_scaling_analysis for log scale in nu
\caption{\textbf{Extracted transport exponents and velocities along the $J_z = \pi$ line.} In top \subfig{a} we see the diffusion exponents $\alpha_\sigma$ extracted from \Cref{fig:variance_drift}\subfig{a}. In the bottom, \subfig{b} reports the integrated speed for the on the time interval $t \in [0,T]$, where $T = N/2$.}
\label{fig:finite_size_scaling_analysis}
\end{figure}

Here we process the realization-averaged spread, $\langle \sigma(t) \rangle$, the drift speed, $\langle \nu(t) \rangle$, and the peak decay, $\langle p_\text{max}(t) \rangle$, to obtain quantities suitable for scaling analysis of transport properties across system sizes $N \in \{14, 16, 18, 20, 22\}$.

To determine the transport exponent $\alpha_\sigma$ from the transient behavior, $\langle \sigma(t) \rangle \sim t^{\alpha_\sigma}$, we focus on intermediate-time regions exhibiting stable log-log behavior (shaded in gray in \Cref{fig:variance_drift}\subfig{a})~\footnote{To refine $\alpha_\sigma$, we perform 25 iterative fits, slightly varying the time window. Each time window is perturbed by a normally distributed factor with mean 1 and $\sigma=0.2$. For each fit, $\langle \sigma(t) \rangle$ is computed from a third of the trajectories (randomly sampled). We then take the logarithm of $\langle \sigma(t) \rangle$ and $t$, fitting the data to $b \ln t + c$, with $\alpha_\sigma$ estimated as the mean of the $b$ coefficients across all runs. This method minimizes sensitivity to both the time window choice and the average $\langle \sigma \rangle$ value. The same procedure is applied to $p_\text{max}$ to estimate $\alpha_p$.}.

The results of the exponent extraction process are displayed in \Cref{fig:finite_size_scaling_analysis}\subfig{a}. Two distinct flat regions emerge: one at $\alpha_\sigma=0$ for very small $J$, and another at $\alpha_\sigma \approx 1/2$ as $J$ increases through the ergodic regime. As discussed, the value $\alpha_\sigma = 1/2$ indicates diffusive transport, though from \Cref{fig:finite_size_scaling_analysis}\subfig{a}, it is unclear if $\alpha_\sigma$ stabilizes at this value as $N$ grows. Larger systems sizes would also be required to determine if the smooth transitions between these flat regions sharpens with increasing system size $N$, which would indicate a clearly defined localized-ergodic crossover. However, our finite-$N$ results are consistent with the static results in \Cref{fig:static-properties}.

\begin{figure}
\includegraphics[width=\columnwidth]{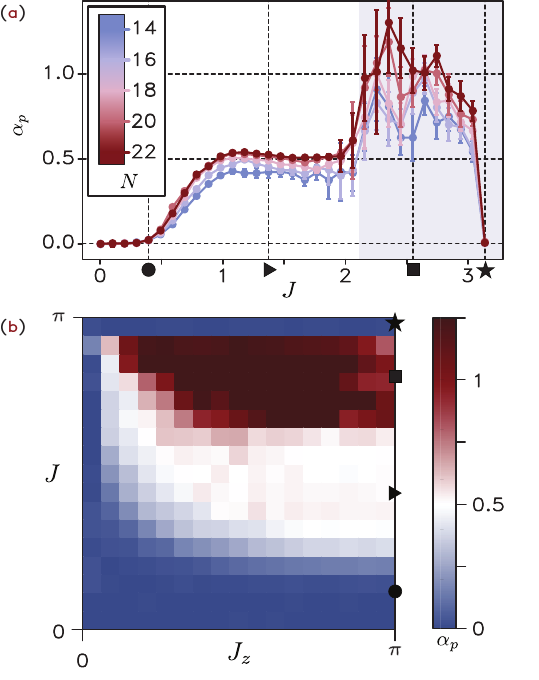}%
\caption{\textbf{Transport exponents derived from peak decay, resulting $J-J_z$ phase diagram.} In panel \subfig{a} we can see a stable area where $\alpha_p \approx 1/2$ and a clear red area above it where the value of $\alpha_p$ also surpasses $1$. Although this last area does not allow to probe transport coefficients it allows to identify where the swappy regime takes place. In panel \subfig{b} the phase diagram of the exponent extracted from the decay of $\langle p_\text{max}(t)\rangle$ at $N=20$ averaged over $30$ trajectories.}
\label{fig:phase_diag}
\end{figure}

In terms of transport properties, this behavior corresponds to a crossover from no transport at very small values of $J$, through an extended sub-diffusive regime, into diffusive transport. 
As $J$ enters the swappy regime (highlighted as a shaded gray region), $\alpha_\sigma$ drops sharply and becomes less stable, as indicated by the wider error bars. This instability is attributed again to the wrapped Normal distribution being an inadequate model for $p_n(t)$ in this regime (see \Cref{fig:variance_drift}\subfig{c}). We later observe a similar instability in $\langle p_\text{max}(t)\rangle$ in \Cref{fig:phase_diag}, wherein the transport exponent fluctuates drastically around unity. Close to the SWAP point $J \to \pi$, the results stabilize as $\alpha_\sigma$ rapidly approaches zero, indicating no spreading of the excitation.

%We remark that in all regimes except ($\blacksquare$) the swappy and ($\bigstar$) near-SWAP regimes, this quantity is well-behaved in this region. In the ($\blacksquare$) swappy regime in particular, the fit region shrinks as early-time rapid transport dominates. We were generally unable to find stable, consistent, fits for values of $\alpha$ in the swappy regime, regardless of fit region.

The results in \Cref{fig:variance_drift}\subfig{b} indicate that the total distance traveled by the excitation increases as $J \to \pi$. It is thus instructive to define a time-averaged drift speed $\overline{\langle \nu \rangle}$ as:
\begin{equation}
    \overline{\langle \nu \rangle} = \frac{1}{T} \int_0^T {\langle \nu(t) \rangle} \, dt
    \label{eq:cumulative}
\end{equation}
which spans from $\overline{\langle \nu \rangle} = 0$ in the localized phase to approximately $\overline{\langle \nu \rangle} \approx 2$ as $J \to \pi$, approaching the SWAP point. In this work, we take $T$ to be the time needed for a SWAP circuit to return an excitation to its original position, $T=N/\nu_\text{typ}$, as discussed in \Cref{supp:typical_drift} of the SM.

The results in \Cref{fig:finite_size_scaling_analysis}\subfig{b} show smooth, stable behavior throughout the $J_z=\pi$ line. We observe that $\overline{\langle\nu\rangle}$ grows linearly across the localized and ergodic regimes, falling exactly to $\overline{\langle\nu\rangle}=0$ in the fully-localized regime ($J\to 0$). Additionally, $\overline{\langle\nu\rangle}$ increases dramatically up to a maximum value of $\overline{\langle\nu\rangle} = 2$ beyond the point $J\approx 2.1$ at which the transport exponent $\alpha_\sigma$ breaks down in \Cref{fig:finite_size_scaling_analysis}\subfig{a}. This behavior remains consistent across all system sizes $N$, but no clear crossover is visible, and a more comprehensive finite-size scaling analysis is needed to extract specific critical values and exponents.
%This aligns with both the local integral of motion (LIOM) picture in MBL, where rapid relaxation occurs, and with the diffusive behavior of the ergodic regime~\cite{Imb-16a, Pal-10a, Ser-13a}. 

In \Cref{fig:variance_drift}\subfig{b}, early-time drift is noticeable in both ergodic and swappy regimes but quickly stabilizes. A sharp rise in $\overline{\langle\nu\rangle}$ in the destabilized $\alpha_\sigma$ region (shaded gray in \Cref{fig:finite_size_scaling_analysis}) suggests coherent late-time information transport. As $N$ increases, we observe a small plateau at $\overline{\langle\nu\rangle}=2$, consistent with \Cref{supp:typical_drift} of the SM, implying that the swappy and near-SWAP regimes remain stable for large $N$, except at ultra-late times.

We analyze the exponent $\alpha_p$ derived from the peak value $\langle p_\text{max}(t)\rangle \sim t^{-\alpha_p}$, using the same method applied for $\langle\sigma(t)\rangle$. As shown in \Cref{fig:phase_diag}\subfig{a}, we find more robust and scale-sensitive results than those in \Cref{fig:finite_size_scaling_analysis}\subfig{a}. The plateaus at $\alpha_p \approx 0$ (full localization) and $\alpha_p \approx 1/2$ (diffusive transport) have stabilized, emerging clearly as scaling limits as $N$ increases. Although results remain somewhat unstable as $J$ approaches the swappy regime, a clearer structure is evident compared to \Cref{fig:finite_size_scaling_analysis}\subfig{a}. Larger system sizes $N$ generally increase $\alpha_p$ in this region, indicating the onset of very rapid thermalization consistent with super-diffusive ($\alpha_p > 1/2$) transport exponents. At $J\to\pi$, $\alpha_p$ nearly vanishes, as expected, since no excitation spreading occurs in this limit. 

%As per our discussion of \Cref{fig:variance_drift}\subfig{c} in \Cref{sec:extracting_transport_properties}, we suggest that these stronger and more stable signatures are caused by the effective subtraction of background noise (realized as multi-modalities in the background data which skews the $\langle \sigma(t)\rangle$) in the extraction of $\langle p_\text{max}(t)\rangle$. 

Finally, we extract the transport exponent $\alpha_p$ from $\langle p_\text{max}(t)\rangle \sim t^{\alpha_p}$ across the entire phase diagram to assess the stability of all regimes away from the $J_z=\pi$ line (and thus the validity of our schematic intuition in \Cref{fig:schematic}). These results are shown for $N=20$ in \Cref{fig:phase_diag}\subfig{b} wherein we identify the four regimes unambiguously: (\ding{108}) $\alpha_p\approx 0$ in the localized regime, (\filledtriangle) $\alpha_p\approx 1/2$ in the ergodic regime, ($\blacksquare$) $\alpha_p > 1/2$ in the swappy regime, and ($\bigstar$) $\alpha_p\approx 0$ in the near-SWAP regime as we approach the DU line. \Cref{fig:phase_diag}\subfig{b} also shows clear crossovers between these regimes. with the localized-ergodic crossover consistent with gap ratio results shown in \Cref{fig:static-properties}\subfig{b}. Surprisingly, away from $J_z=\pi$ we find exponents consistent with ballistic $\alpha_p \approx 1$ transport or super-ballistic $\alpha_p > 1$ transport; though the connection between the relaxation of local observables to true transport properties is tenuous, and the nature of transport via e.g., conductance measurements is a topic worth interrogating in future research. We discuss these large exponents again in the heuristic analysis of the swappy regime in the next section.

\paragraph{Prethermalization and the swappy regime.}
\label{sec:prethermal}

These results collectively establish the existence of a swappy regime as a distinct feature unique to discrete-time many-body systems, and located between the ergodic regime and DU line. Here we interrogate the stability and nature of this regime. 

\begin{figure}
    \centering
    \includegraphics[width=\linewidth]{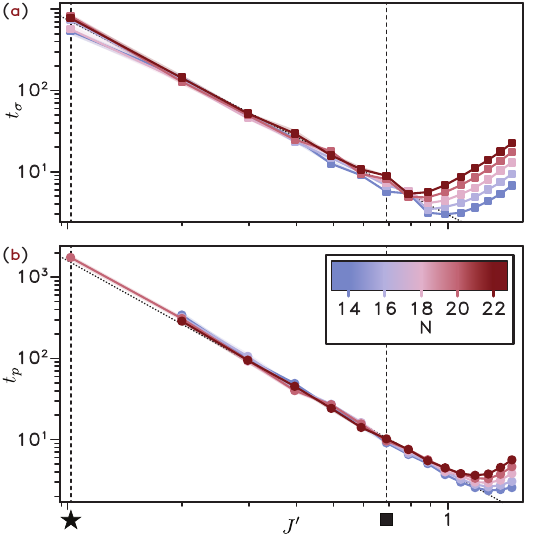}
    \caption{\textbf{Prethermal scaling near the SWAP point.} Values are extracted from $\sigma(t)$, as $t_\sigma = \inf\{t \mid \sigma(t) > N/30 + 1.25\}$ (\subfig{a}), and from $p_\text{max}(t)$, as $t_p = \inf\{t \mid p_\text{max}(t) < 2.6/N\}$ (\subfig{b}). The dotted lines are $t \propto (J')^{-2.4}$ and $t \propto (J')^{-2.6}$, respectively.}
    \label{fig:prethermal}
\end{figure}

Based on our previous results, we expect that in the swappy and near-SWAP regimes, the system will eventually fully thermalize at very late times. However, in this section, we argue that these regimes are characterized by prethermal behavior, i.e., that the time needed for thermalization diverges as we approach the integrable SWAP point. To verify this interpretation of the swappy regime, we present a numerical investigation of the time taken for the system to thermalize. We estimate thermalization times via two operational definitions: $t_\sigma$, the first time at which $\langle\sigma(t)\rangle$ exceeds a threshold, and $t_p$, the first time at which $\langle p_\text{max}(t)\rangle$ falls below a threshold. We expect these to give similar results. We examine the behavior of the thermalization timescale as a function of the parameter $J^\prime = \pi - J$, which represents the deviation of $J$ from the integrable SWAP point at $J = \pi$, and at which we expect the thermalization time to diverge. Our numerical results are plotted in \Cref{fig:prethermal}\subfig{a} and \subfig{b}. We see that, as $J'$ decreases to zero (i.e., as we approach the SWAP point) the thermalization times diverge as $t_p \propto (J')^{-2.6}$ and $t_\sigma \propto (J')^{-2.4}$. Both exhibit a power-law scaling with an exponent $\approx 2$ that is expected generically for prethermalization due to proximity to an integrable point~\cite{Mor-22a}. This implies the existence of an extended prethermal swappy regime. We note that the precise numerical values of the power-law exponents for the thermalization times are contingent on our ad-hoc choice of the threshold values for $\langle p_{\text{max}}(t) \rangle$ and $\langle \sigma(t) \rangle$ that determine our thermalization condition. However, we have checked that the exponents are $\approx 2$ for a broad range of threshold values.

%Moreover, this result indicates that thermalization occurs after timescales $t J^{\prime 2} \sim 1$, which is consistent with the form of \Cref{eqn:jprime-perturbed} and our associated discussion of the underlying mechanism as essentially a classical Markov chain. Essentially, at times in excess of $t J^{\prime 2} \sim 1$ the $\mathcal{O}(J^{\prime 2})$ terms in \Cref{eqn:jprime-perturbed} become non-trivial, and the local dynamics becomes more complicated than simple applications of SWAP and identity gates, leading to a breakdown of the probabilistic Markov chain mechanism, and thermalization. Moreover, this result exhibits clear scale-invariance up to the point $\pi-J \sim 1$, consistent with the ergodic-swappy crossover seen in \Cref{fig:finite_size_scaling_analysis} and \Cref{fig:phase_diag}\subfig{a}. Altogether these results establish the swappy regime as a prethermal, dynamical regime.

% These results, taken in concert, strongly evidence the existence of a distinct, stable, prethermal swappy regime; unique to discrete many-body systems.
% Since both the ranges of variability of $\sigma$ and $p_{\text{max}}$ scale with the system size $N$, the threshold values for these quantities will also depend on $N$, as discussed in \Cref{fig:prethermal}. Additionally, as we approach the SWAP point, these times increase, following a power law with an exponent of approximately $-2$ and almost no dependence on $N$. This behavior suggests that $t \sim (1/\Delta J)^{2}$, where $\Delta J = \pi - J$, consistent with expectations in a prethermal regime~\cite{Mor-18a, Yan-23a}.

\section{Discussion}
\label{sec:conclusion}

Our results address transport across various phases of a highly generic $\mathrm U(1)$-symmetric disordered Floquet model, revealing a wide array of phenomenological behaviors. This includes more conventional localized and ergodic regimes, exhibiting sub-diffusive and diffusive transport respectively; and the existence of a ``swappy'' regime unique to discrete-time quantum systems, in which excitations propagate ballistically. To enable this investigation, we study the model from both static and dynamic perspectives, by focusing in the zero magnetization subsector. The former is pursued by means of POLFED, adapted to work in a given magnetization subsector by applying the unitary gate-by-gate. We have developed and deployed the first circular moment (see \Cref{fig:spirals} and discussion thereof): a quantity defined for periodic systems that directly encodes their transport properties, and which maps their aggregated dynamics onto simple yet striking two-dimensional diagrams. This quantity provides a visual and intuitive way to understand transport across different regimes in periodic systems; and, as it is singularly composed of local $Z$-basis expectation values, is highly amenable to experimental implementation; paving the way for the near-future experimental analysis of transport in digital matter. Of this quantity we analyze the decay, connected to the spread of the initial excitation, and speed of its phase, from which we define a notion of drift. Our drift might be linked to the concept of entanglement speed~\cite{Zho-22a, Ram-24a}, particularly in the near-SWAP regime, where it aligns with the characteristics of entanglement speed in DU circuits.

We emphasize that the dynamics studied here are not obtained from a simple Trotterization of a continuous-time disordered XXZ Hamiltonian because the disorder strength in our model does not scale with the kicking parameters. Physically, our protocol corresponds to strong static disorder that is periodically interrupted by finite unitary kicks. The non-scaling of disorder produces interference effects and a nearby discrete-time integrable point that do not survive the usual continuous-time limit; these give rise to qualitatively new transport behavior, in particular the ``swappy'' regime, which we identify and characterize below. Moreover, the circular statistical moments we use are directly measurable in current quantum-simulator experiments, providing an immediate route to test the predicted Floquet-specific phenomena.

Our results, taken in concert, demonstrate the existence of a swappy regime which is (i) distinct: exhibiting a clear ergodic-swappy crossover in the vicinity of $J\sim 2.1$ for all investigated dynamical quantities of \Cref{sec:transport_properties}, (ii) dynamical: as it is absent in the static analyses of \Cref{sec:spectral_properties}, (iii) prethermal: exhibiting a failure to thermalize until late timescales which grow as a power law in $t \sim (\pi-J)^{-2}$ (see \Cref{sec:prethermal}), (iv) stable to perturbations in $J$ and $J_z$ (see \Cref{fig:phase_diag}\subfig{b} and discussion thereof), (v) stable as a function of system size (see, e.g., \Cref{fig:finite_size_scaling_analysis} and \Cref{fig:prethermal}), and (vi) unique to discrete-time Floquet systems, appearing in the vicinity of the DU line at $J=\pi$, which has no counterpart in continuous models. We also provide a rough outline of a potential semi-classical mechanism underlying this regime; though we defer a detailed analysis of this mechanism to future study.

\section{Methods}
\label{supp:methods}

\subsection{POLFED}
\label{supp:polfed}

The entanglement entropy values we compute are derived from the mean of $\min({d}/10, 750)$ eigenvectors. To acquire these eigenvectors, we employ polynomial filtered diagonalization (POLFED)~\cite{Lui-21a}, a spectral transformation akin to the shift-invert method, tailored for unitary matrices. 
The Floquet operator $\hat{\mathbb{U}}$ is redefined as the operator
\begin{equation}
\hat{\mathbb{K}} = \sum_{k=0}^{K} e^{-i k \phi_\text{tgt}} \hat{\mathbb{U}}^k
\label{eq:polfed}
\end{equation}
This new operator is non-unitary yet retains the eigenvectors of the original Floquet operator. Moreover, the eigenphases of $\hat{\mathbb{U}}$ within the interval delimited by $(\phi_\text{tgt} \pm {2\pi}/{K})$ are now transformed such that their absolute values exceed $1$, while the others are less than $1$.
This enhances the convergence properties of the Arnoldi method, which we employ to derive the eigenphases and eigenvectors of $\hat{\mathbb{U}}$. In this study, we consistently selected $\phi_\text{tgt}=0$. For the parameter $K$, we adopted the recommendation from Ref.~\cite{Sie-20b, Lui-21a}, setting $K$ to $0.4\ d / \min(d/10, 750)$.\\

\subsection{Efficient time evolution of a state by a random circuit}
\label{supp:time_evolution_method}

The numerical method we developed circumvents the need to construct the entire $\hat{\mathbb{U}}$ matrix. Instead, we apply each gate directly to the state vector, leveraging the fact that both static and dynamic analyzes necessitate the application of the Floquet operator to a state. This allows the memory cost of diagonalizing $\hat{\mathbb{U}}$ to be $O(d)$ and the number of operations to be $O(Ld)$, similar to what pointed out in Ref.~\cite{Mor-22a}.\\
Our approach involves recording the action of a gate on the first two spins of a state $\ket{\psi}$. If a gate acts on spins $a$ and $a+1$, we shift the system by $-a$ sites, apply the gate to the first two spins, and then shift the system back to its original configuration.\\
To implement this, we require two key components. First, we must determine the value of the first two spins for each element of $\ket{\psi}$. The state is represented as a ${d}$-vector:
\begin{equation}
    \ket{\psi} = \sum_i c_i \ket{i}
\end{equation}
where the $i$-th element corresponds to the $\ket{i}$ state. Hence, we need to identify which segments of the ${d}$-vector have the first and second spins up, namely which of the $c_i$ are associated with a $\ket{i}$ with first and second spin up.\\
The second ingredient involves the transformation $f_{(1)}$, which maps a state from sites $0, \dots, N-1$ to the corresponding state on sites $N-1, 0, \dots, N-2$. By recursively applying $f_{(1)}$, we can achieve the required shift, i.e., $f_{(a)}=f_{(1)}^a$. This strategy offers a significant memory advantage, as it only necessitates storing vectors of size ${d}$. Although this comes with a cost in the number of operations---each gate of $\hat{\mathbb{U}}$ or $\hat{\mathbb{K}}$ must be applied sequentially---the memory efficiency of this method allows for better parallel computation of multiple trajectories, which is crucial for disorder averaging.\\

\subsection{Efficient computation of the entropy in a magnetization subsector}
\label{supp:entropy_computation}

To compute the entropy, we can reduce the computation to different magnetization sub-sectors. 
Specifically, the entanglement entropy, given by \Cref{eq:entanglement_entropy}, can be calculated by reshaping the state $\ket{\psi}$ into a ${d}_{N_A} \times {d}_{N_B}$ matrix and extracting its Schmidt coefficients ${s_i}$ via Singular Value Decomposition (SVD). 
The von Neumann entropy of $\ket{\psi}$ is then
\begin{equation}
    S(\ket{\psi}) = - \sum_i s_i^2 \ln s_i^2\,.
\end{equation}
Among these steps, the SVD is the most computationally intensive. However, we can exploit the structure of the system further. The space ${d}_{N_A}$ can be decomposed as
\begin{equation}
    \mathcal{H}^{(M)}_A = \bigoplus_{M_A=-N/2}^{N/2}  \mathcal{H}^{(M)}_{A, (M_A)}.
\end{equation}
Given that $\ket{\psi}$ resides in $\mathcal{H}^{(M)}$ with $M = 0$, if the $A$ part has magnetization $M_A$, the $B$ part will have magnetization $M_B = M - M_A$. Consequently, for each $M_A$ value we can isolate the component of $\ket{\psi}$ in $\mathcal{H}^{(M)}_{A, (M_A)}$, reshape it into a ${d}^{(M)}_{N_A, (M_A)} \times {d}^{(M)}_{N_B, (M_B)}$ matrix, and compute the SVD on this smaller matrix. 
In other terms, we split the zero-magnetization sub-sector as
\begin{equation}
    \mathcal{H}^{(M)} = \bigoplus_{M_A=-N/2}^{N/2} \mathcal{H}^{(M)}_{A, (M_A)} \otimes \mathcal{H}^{(M)}_{B, (M-M_A)}.
\end{equation}
The total entanglement entropy of the state is then given by
\begin{equation}
    S(\ket{\psi}) = - \sum_{M_A} \sum_{i\in M_A} s_i^2 \ln s_i^2.
\end{equation} 
By sidestepping the full $d$-state vector's SVD, the process of extracting entanglement entropy becomes more efficient and significantly more manageable.\\

\subsection{Equilibrium diagnostics and their interpretation}
\label{supp:infinite_time_diagnostics}

We briefly justify why the diagnostics used in the analysis of the spectral properties of the model, the mean eigenstate entanglement entropy and the mean gap ratio, characterize equilibrium behavior in the infinite-time limit.

For a closed many-body system, an initial state $|\psi(0)\rangle=\sum_n c_n|E_n\rangle$ dephases at late times, and observables are governed by the diagonal ensemble
\begin{equation}
\overline{\langle \psi(0) | \hat{O}(t)| \psi(0) \rangle}=\sum_n |c_n|^2 \bra{E_n} \hat{O} \ket{E_n},
\end{equation}
where the time average is defined in \Cref{eq:cumulative} with $T\to\infty$, such that infinite-time averages are fixed by eigenstate expectation values. ETH gives a statistical (RMT-like) form for matrix elements, implying that typical eigenstates at a given energy have thermal reduced states~\cite{Deutsch1991,Srednicki1994}. The Page result provides the quantitative benchmark for entanglement of a random pure state (and its variants for fixed-charge sectors), hence, if eigenstates are effectively random within the relevant symmetry/charge sector, the mean eigenstate von Neumann entropy should approach the corresponding Page value~\cite{Pag-93a,Bianchi2019}. In practice, we therefore compare the numerically obtained mean eigenstate entanglement to the Page value computed in the fixed-magnetization subsector (see main text and \Cref{eq:right_page_value} above). Deviations from the Page value indicate either finite-size corrections or a breakdown of eigenvector randomness (e.g., integrability, many-body localization, or scarred eigenstates).

The mean gap ratio \Cref{eq:gap_ratio} is a purely spectral statistic that distinguishes Wigner--Dyson (level repulsion) from Poisson (level clustering) statistics. Wigner--Dyson statistics signal RMT-like eigenvectors~\cite{Bohigas1984}, which underpins the randomness assumption in ETH and the Page-like entanglement of typical eigenstates~\cite{Pag-93a}. Thus, the gap ratio (spectral correlations) and the mean eigenstate entanglement (eigenvector/statistical properties) are complementary diagnostics of whether the system's infinite-time behavior is thermal (ergodic) or non-thermal.\\

\section{Declaration statements}

\subsection{Data availability}

All data is available at \url{https://github.com/alessum/swappy}.

\subsection{Code availability}

The data and the code generated throughout this project can be found at \url{https://github.com/alessum/swappy}.

\subsection{Acknowledgements}
The simulations were performed on the Luxembourg national supercomputer MeluXina, and the authors thank the LuxProvide teams for their expert support. The authors thank members of the QuSys group for discussions throughout the project. A.S. and J.G. acknowledge Iman Marvian for insightful conversations. A.N.-K. acknowledges support through the SFI-IRC Postdoctoral Fellowship GOIPD/2025/1504. S.D. acknowledges support through the SFI-IRC Pathway Grant 22/PATH-S/10812. J.G. is supported by a SFI-Royal Society University Research Fellowship and is grateful to IBM Ireland and for generous financial support. A.S., A.N.-K., and J.G. gratefully acknowledge the financial support provided by Microsoft Ireland for their research.

\subsection{Author contributions}
All authors contributed equally to the conception and development of the project. A.S., S.D., and A.N.-K. carried out the computational aspects of the work. A.S. wrote the initial manuscript and served as corresponding author, with revisions made by  A.S., A.N.-K., and S.D. throughout the peer review process. All authors, A.S., A.N.-K., S.D., and J.G., have contributed to, read, and approved the final manuscript.

\subsection{Competing interests}
The authors declare no competing interests.

\section*{References}
\bibliography{refs}

% ---------- Supplementary Materials ----------
\clearpage
\onecolumngrid
\begin{center}
{\large\bfseries Supplementary Materials: Anomalous transport in U(1)-symmetric quantum circuits\par}
\vspace{0.8em}
Alessandro Summer\textsuperscript{1,2,*},
Alexander Nico-Katz\textsuperscript{1,2,\(\dagger\)},
Shane Dooley\textsuperscript{3,\(\ddagger\)}, and
John Goold\textsuperscript{1,2,4,\(\mathsection\)}\\
\textit{\textsuperscript{1}School of Physics, Trinity College Dublin, Dublin 2, Ireland}\\
\textit{\textsuperscript{2}Trinity Quantum Alliance, Unit 16, Trinity Technology and Enterprise Centre, Pearse Street, D02 YN67, Dublin 2, Ireland}\\
\textit{\textsuperscript{3}Dublin Institute for Advanced Studies, School of Theoretical Physics, 10 Burlington road, Dublin, D04 C932, Ireland}\\
\textit{\textsuperscript{4}Algorithmiq Ltd, Kanavakatu 3C 00160, Helsinki, Finland}\\
(Dated: November 20, 2025)
\vspace{0.5em}\\
{\footnotesize
\textsuperscript{*}\href{mailto:summera@tcd.ie}{summera@tcd.ie}\quad
\textsuperscript{\(\dagger\)}\href{mailto:nicokatz@tcd.ie}{nicokatz@tcd.ie}\quad
\textsuperscript{\(\ddagger\)}\href{mailto:dooleys@tcd.ie}{dooleys@tcd.ie}\quad
\textsuperscript{\(\mathsection\)}\href{mailto:gooldj@tcd.ie}{gooldj@tcd.ie}}
\end{center}
\vspace{2em}

% Reset counters and redefine numbering
\setcounter{section}{0}
\renewcommand\thesection{S~\arabic{section}}
\renewcommand\thesubsection{\thesection\Alph{subsection}}
\setcounter{equation}{0}
\renewcommand\theequation{S\arabic{equation}}
\setcounter{figure}{0}
\renewcommand\thefigure{S\arabic{figure}}

\section{Typical drift}
\label{supp:typical_drift}

\begin{figure}
    \centering
    \includegraphics[width=0.7\linewidth]{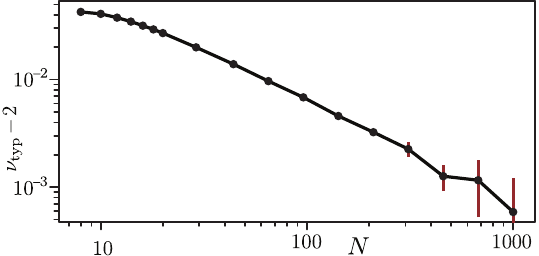}
    \caption{
    \textbf{Comparison of typical brickwork drift as a function of system size.} For $N\leq 10$ the value is exact, by studying all the $N!$ permutations. We sample $10^7$ trajectories for $N\leq 20$, then $10^7/3$ for $N<50$, $10^6$ for $N<250$, $10^6/3$ for $N<500$, otherwise $10^5$.}
    \label{fig:typ_drift}
\end{figure}

The random permutation of gates enables the restoration of translational invariance by averaging over different permutations. This averaged circuit will have some averaged properties such as an average drift. To analyze this, we can consider a circuit composed exclusively of generalized SWAP gates (i.e., $J=\pi$, as from \Cref{eq:U_swap}) and investigate the resultant circuit after averaging over these permutations. Let us focus on the scenario with same initial state as \Cref{eq:initial-state}, namely with an excitation located at the middle site of the chain. We can define the drift as an effective speed, calculated as the number of sites traversed to return to the initial one ($N$ sites) divided by a time. This time corresponds to the number of gates applied, normalized over $N$ (i.e., the number of gates per Floquet cycle).\\
By studying this, we observe that most efficient circuits for returning the excitation to its starting site are the two staircase circuits that begin with the gate $\hat {U}_{N/2,N/2+1}$ (or $\hat {U}_{N/2-1,N/2}$), employing the permutation $P: i \mapsto N/2 - 1 + i$ (or $P: i \mapsto N/2 - i$). These configurations yield a drift value of $\nu=N$. Conversely, when the staircase geometry is reversed, using permutations like $P: i \mapsto N/2 - 2 + i$ or $P: i \mapsto N/2 + 1 - i$, the drift value drops to $N^2/(N^2-2N+2)$, which tends to 1 as $N$ grows.\\
Calculating the drift precisely involves evaluating all   $N!$ possible permutations, and we could not express this in a closed form. However, by sampling a large number of permutations, we can accurately estimate the average drift very accurately. \\
Our findings reveal that for $N \geq 8$, the average drift decreases monotonically towards $2$, corresponding to the value associated with brickwork circuits. Since brickwork circuits are the most commonly obtained configuration under random $P$, we define this average drift as the typical drift, denoted by $\nu_\text{typ}$. For instance, at $N = 20$, after sampling $10^7$ trajectories, the average drift was found to be $\nu_\text{typ} \approx 2.027(6)$. This power-law-like trend approaching $2$ is presented in \Cref{fig:typ_drift}.

\section{Reduction of four phases to two plus the Peierls phase in {U(1)}-symmetric gates}
\label{supp:pauli_to_peierls}
Although the gate $\hat{U}_{n,n+1}$, defined in \Cref{eq:two-qubit-gate}, remains the most general $\mathrm U(1)$-symmetric gate, we preferred, for the purposes of our numerical method, to precompute $\hat{H}_{n,n+1}$ without phases and then add them individually gate by gate.

The initial gate we employ is
\begin{equation}
\begin{split}
\hat{V}_{n,n+1}=&e^{-i\theta_1^{(n)}/2\hat{\sigma}_n^z} e^{-i\theta_2^{(n)}/2\hat{\sigma}_{n+1}^z} \\
& e^{-iJ/4\,\hat{\sigma}_n^x\hat{\sigma}_{n+1}^x} e^{-iJ/4\,\hat{\sigma}_n^y\hat{\sigma}_{n+1}^y} e^{-iJ_z/4\,\hat{\sigma}_n^z\hat{\sigma}_{n+1}^z} \\
& e^{-i\theta_3^{(n)}/2\hat{\sigma}_n^z} e^{-i\theta_4^{(n)}/2\hat{\sigma}_{n+1}^z}
    \label{eq:two-qubit-gate-original}
\end{split}
\end{equation}
where the Pauli operators are swapped out for spin operators:
\begin{align}
    \hat{S}^z = \hat{\sigma}^z/ 2 \qquad \hat{S}^\pm = \hat{\sigma}^\pm.
    \label{eq:Pauli_to_spin_op}
\end{align}
This substitution yields the expression:
\begin{equation}
\begin{split}
\hat{V}_{n,n+1}=&e^{-iJ_z\,\hat{S}_n^z\hat{S}_{n+1}^z} e^{-i\theta_1^{(n)}\hat{S}_n^z} e^{-i\theta_2^{(n)}\hat{S}_{n+1}^z} \\
& e^{-iJ/2\,\left(\hat{S}_n^+\hat{S}_{n+1}^-+\hat{S}_n^-\hat{S}_{n+1}^+\right)}\\
& e^{-i\theta_3^{(n)}\hat{S}_n^z} e^{-i\theta_4^{(n)}\hat{S}_{n+1}^z}.\\
\end{split}
\end{equation}
Due to the commutativity of the two-body terms, this simplifies to:
\begin{equation}
\begin{split}
\hat{V}_{n,n+1}=&e^{-iJ_z\,\hat{S}_n^z\hat{S}_{n+1}^z} e^{-i\theta_1^{(n)}\hat{S}_n^z-i\theta_2^{(n)}\hat{S}_{n+1}^z} \\
& e^{-iJ/2\,(\hat{S}_n^+\hat{S}_{n+1}^-+\hat{S}_n^-\hat{S}_{n+1}^+)}\\
& e^{i\theta_1^{(n)}\hat{S}_n^z+ i\theta_2^{(n)}\hat{S}_{n+1}^z}\\
& e^{-i(\theta_1^{(n)}+\theta_3^{(n)})\hat{S}_n^z-i(\theta_2^{(n)}+\theta_4^{(n)})\hat{S}_{n+1}^z}\\
\end{split}
\end{equation}
where the $\hat{S}^\pm$ operators, flanked by opposite phases, generate the Peierls phase, condensing the notation into:
\begin{equation}
\begin{split}
\hat{V}_{n,n+1}=&e^{-iJ_z\,\hat{S}_n^z\hat{S}_{n+1}^z} \\
& e^{-iJ/2\,\left(\hat{S}_n^+\hat{S}_{n+1}^-e^{i(\theta_1^{(n)} - \theta_2^{(n)})}+\hat{S}_n^-\hat{S}_{n+1}^+e^{-i(\theta_1^{(n)} - \theta_2^{(n)})}\right)}\\
& e^{-i(\theta_1^{(n)}+\theta_3^{(n)})\hat{S}_n^z-i(\theta_2^{(n)}+\theta_4^{(n)})\hat{S}_{n+1}^z}\\
=& \hat{U}_{n,n+1}\\
\end{split}
\end{equation}
where $\hat{U}_{n,n+1}$'s parameters are: 
\begin{equation}
    \phi_n = \theta_1^{(n)} - \theta_2^{(n)}\!\!\!,\;\; h_n=\theta_1^{(n)}+\theta_3^{(n)}\!\!\!,\;\text{and }h'_n=\theta_2^{(n)}+\theta_4^{(n)}\!\!\!.
\end{equation}
Owing to the $2\pi$ periodicity of $\phi_n, h_n,$ and $h'_n$, the sum or difference of the uniformly distributed $\theta_k^{(n)}$ within $[-\pi, \pi]$ ensures that the resulting parameters are uniformly distributed in $[-\pi, \pi]$ as well.

This eventually proves that the black gates of \Cref{fig:schematic}\subfig{a} can be decomposed either as \Cref{eq:two-qubit-gate} or \Cref{eq:two-qubit-gate-original}:
\begin{equation}
\begin{array}{rcl}
\includegraphics[width=.85\linewidth]{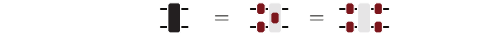}
\end{array}
\label{eq:image_equation}
\end{equation}

\section{Spectral Properties and Dynamical Correlations via Transfer Matrix
\label{supp:transfer_matrix}
}

The computation of dynamical correlations in dual-unitary circuits is greatly simplified by the transfer matrix structure. 
Along the light cone, correlation functions
\begin{equation}
    C_{n,n'}(t)=\Tr[\hat{S}_{n}^z(t)\hat{S}^z_{n'}]
\end{equation}
can be expressed in terms of the action of the transfer matrix
\begin{equation}
    \mathcal{M}[\rho] = \Tr_1\left[\hat{U}_{0,1} (\hat{\rho} \otimes \hat{\mathbb{I}}_2) \hat{U}_{0,1}^\dagger\right],
\end{equation}
which encodes the effective evolution of a single qubit under the two-qubit unitary gate. 
This allows us to rewrite nearest-neighbor correlation functions as
\begin{equation}
    C_{0,1}(t)=\Tr[\hat{S}_{0}^z(t)\hat{S}^z_{1}]=\Tr[\mathcal{M}[\ket{\!\uparrow}\bra{\uparrow\!}_0]\hat{S}^z_{1}].
\end{equation}
As shown in Ref.~\cite{Ber-19a} of the main text, for dual-unitary circuits this is the unique non-trivial correlator that survives the maximal entanglement, making the light-cone correlations the fundamental observable to study.

The spectral properties of $\mathcal{M}$ determine the long-time behavior of dynamical correlations. 
Diagonalizing the transfer matrix yields four eigenvalues
\begin{align}
\lambda_0 &= 1 \quad \text{(identity)}, \\
\lambda_z &= \frac{1 - \cos J}{2} \quad \text{(population)}, \\
\lambda_\pm &= \pm \sin\frac{J}{2}\sin\frac{J^z}{2} \quad \text{(coherence)},
\end{align}
with corresponding eigenmodes
\begin{align}
\hat{r}_0 &= \hat{\mathbb{I}}_2, \\
\hat{r}_z &= \hat{S}^z, \\
\hat{r}_\pm &= \begin{pmatrix}0& e^{i(h_{1} - \varphi_0)}\\ e^{-i(h_{1} - \varphi_0)} & 0 \end{pmatrix}.
\end{align}
Crucially, the eigenvalue spectrum is independent of the random phases $h_0$, $h_1$, and $\varphi_0$, which appear only in the eigenmodes $\hat{r}_\pm$. 
This phase-independence allows exact computation of correlation functions on the light cone, as long as finite size effects kick in.

\begin{figure}
    \centering
    \includegraphics[width=0.7\linewidth]{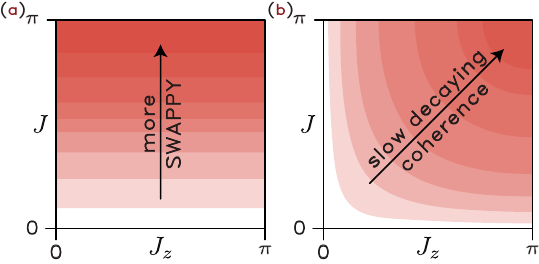}
    \caption{\textbf{Properties of a single U(1) gate extracted from the transfer matrix eigenvalues.} Panel \subfig{a} shows the value of the $\lambda_z$ as a function of $J$ which determines population mixing. Panel \subfig{b} shows absolute values of the $\lambda_\pm$ increasing towards the SWAP point.}
    \label{fig:transfer_matrix}
\end{figure}

The entire dual-unitary family with $J = \pi$ displays non-ergodic dynamics according to the definition of Ref.~\cite{Ber-19a} of the main text, characterized by the presence of at least two independent conserved quantities. 
As the Floquet unitary approaches the DU line the circuit, it inherits exact local conservation laws of dual-unitary models, such (near-)conserved operators fragment the spectrum into overlapping symmetry sectors and therefore distort raw level-spacing statistics unless those symmetry sectors are first resolved~\cite{Hol-25a}. This connects with the instability in the spectral quantities observed in \Cref{sec:spectral_properties}.

\section{Discretized Wrapped Normal distribution}
\label{supp:discretized-results}

\begin{figure}
    \centering
    \includegraphics[width=0.7\linewidth]{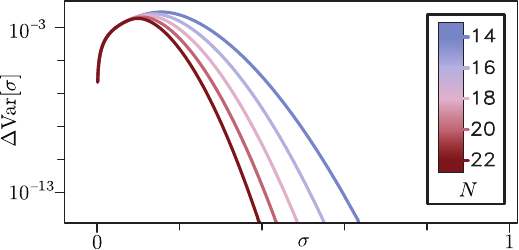}
    \caption{\textbf{Difference between the circular variance of the discrete and continuous wrapped normal distributions.} Where $\Delta \text{Var}[\sigma]$ is defined as the difference in variance between the continuous and discrete wrapped normal distributions.}
    \label{fig:discrete_WN_diff}
\end{figure}

As discussed in \Cref{sec:quasi-prob}, to quantify the rate of spreading, we compare the distribution of $p_n(t)$ with a wrapped Normal distribution. Since the wrapped Normal distribution has a continuous support, a more rigorous approach might be to discretize it. This discrete version of the wrapped Normal distribution would be defined as:
\begin{equation}
    \tilde{q}_n(\sigma) = \sum_{k=-\infty}^{\infty} e^{-(\theta_n + 2\pi k)^2 / 2\sigma^2},
\end{equation}
with the normalized form given by
\begin{equation}
    q_n(\sigma) = \frac{\tilde{q}_n(\sigma)}{\sum_n \tilde{q}_n(\sigma)},
    \label{eq:discrete_wrapped_normal}
\end{equation}
which introduces a more complex procedure for extracting $\sigma$, as there is no closed-form relationship between $\sigma$ and $Var[\sigma]$. However, for already small values of $N$, the differences between the continuous and discrete distributions are minimal, and the added complexity does not significantly improve accuracy. 

In \Cref{fig:discrete_WN_diff}, we compare the variances of the continuous and discrete wrapped Normal distributions for different values of $N$. Our results indicate that, given the number of qubits in our simulations, the two distributions closely align, showing that the continuous approximation is sufficiently accurate.

\begin{figure}
    \centering
    \includegraphics[width=0.7\linewidth]{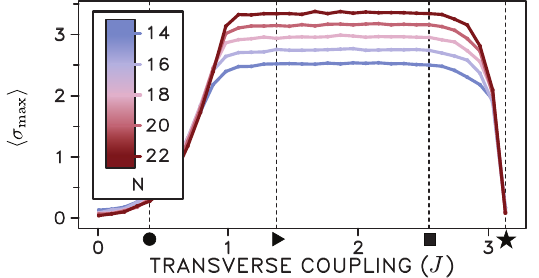}
    \caption{\textbf{Extensive scaling of maximum values of the circular variance along the $J_z=\pi$ line.} Properties of the disorder-averaged maximum value of the circular variance $\langle\sigma_\mathrm{max}\rangle$ as a function of system size $N$ and $J$ along the $J_z=\pi$ line. Shows extensive behaviour (linear growth as a function of $N$) away from the localized and near-SWAP regions.}
    \label{fig:max_sigma}
\end{figure}

\section{Finite size scaling of saturation values of the spread parameter}
\label{supp:finite-size-plateau}

In \Cref{sec:scaling-analysis} of the main text, we observed a distinct plateau in the late-time behavior of $\langle\sigma(t)\rangle$ as systems reach thermalization. This observation might seem counterintuitive, given that in a fully thermalized state, we would expect a uniform distribution with $|R(t)| \to 0$, implying $\sigma(t) \to \infty$ according to \Cref{eq:wrapped_normal}. 

Here, we provide straightforward numerical evidence indicating that this plateau is a finite-size effect. Possible explanations for this effect include: (a) the wrapped Normal distribution only accurately represents $p_n(t)$ in the continuum limit $N \to \infty$, or (b) finite-dimensional systems lack the ability to fully self-thermalize. In \Cref{fig:max_sigma}, we display the maximum value of $\sigma$ over the first $T = 1000$ time steps for various values of $N$. For values of $J$ where the system thermalizes within this timeframe, $\langle \sigma_\text{max}\rangle$ reaches a plateau that grows steadily with $N$.

\section{Expanded Transport}
\label{supp:expanded-transport}

Transport coefficients and exponents in many-body physics are typically derived from linear response theory~\cite{Ber-21a}. 
In this Supplementary Note we briefly review how the magnetization profile $M_n(t)$ for our particular choice of the initial random state $\hat{P}^\uparrow_{N/2}\ket{\psi_\text{rand}}$, an in particular the resulting semiprobability $p_n(t)$, corresponds to the two-point correlation function of the spin $z$-operator $\hat{S}^z_n$.\\
We divide our discussion into three steps. 
First, we illustrate how the spin profile of our initial state aligns with the correlation function of $\hat{S}_n^z$ within the complete Hilbert space. 
Second, we demonstrate how this can be tailored to the $M$-sub-sector. 
Finally, we connect this framework to typical random states, thereby fully linking our discussion to the definition of $M_n(t)$ of \Cref{sec:numerical_magnetization_profile}.

\subsection{Dynamical correlation function in the full Hilbert space}
\label{supp:typical_state}

These coefficients are extracted from spatiotemporal correlations, which, in systems that conserve the total spin component $\hat{S}^z_\text{tot}$, are defined as follows:
\begin{equation}
    C_{n,n'}(t)=\Tr[\hat{S}_{n}^z(t)\hat{S}^z_{n'}].
    \label{eq:corr_function}
\end{equation}
First we observe that the unpropagated spin operator can be decomposed as:
\begin{align}
\hat{S}_{n^\prime}^z &= \hat{\mathbb{I}}_2^{\otimes N-1} \otimes \left(\ket{\!\uparrow\,}\bra{\,\uparrow\!} - \hat{\mathbb{I}}_2/2 \right)_{n^\prime} \\ 
&= \hat{P}^\uparrow_{n^\prime} - \hat{\mathbb{I}}_2^{\otimes N}/2 \label{eqn:projector-alternate-spin-z}
\end{align}
where $\hat{\mathbb{I}}_2$ is the local two-dimensional identity, the subscript $n^\prime$ denotes which local Hilbert space the relevant local spin-z operator must be inserted into, and where the projector $\hat{P}^{\uparrow}_{n^\prime} = \hat{\mathbb{I}}_2^{\otimes N - 1} \otimes \ket{\!\uparrow\,}\bra{\,\uparrow\!}_{n^\prime}$ projects the $n^\prime$-th site onto the spin-up state. Direct substitution of \Cref{eqn:projector-alternate-spin-z} into the correlation function yields 
\begin{equation}\label{eqn:correlation-function-projected}
    C_{n,n'}(t)= \Tr\left[\hat{S}_n^z(t) \hat{P}^{\uparrow}_{n^\prime} \right]-\Tr[\hat{S}_n^z(t)]/2.
\end{equation}
wherein the second term reduces to zero by simple observation that spin operators are traceless $\Tr[\hat{S}_n^z(t)] = 0$. The first term intuitively corresponds to the simple expectation value of the spin-z operator on the $n$-th site at time $t$, conditioned on the initial state being a maximally mixed state with the $n^\prime$-th site projected into the spin-up state.

This intuition becomes even more developed when one considers the stochastic trace estimation approach to actual numerical computation of \Cref{eqn:correlation-function-projected}. In practice, explicit calculation of the trace is computationally expensive, with a number of operations that na\"ively scales as $D^2$ (where $D$ is the dimensionality of the total system). It is therefore commonly computed stochastically ~\cite{Ski-98a, Wei-06a}:
\begin{equation}\label{eqn:stochastic-trace-estimation}
    C_{n,n'}(t) \simeq \frac{1}{R} \sum_{r=1}^{R} \bra{\psi_r} \hat{S}_n^z(t) \hat{P}^\uparrow_{n^\prime} \ket{\psi_r}.
\end{equation}
where $\ket{\psi_r}$ are random states, generated following the method of~\cite{Wei-06a}. The accuracy of this approximation follows central limiting behavior, scaling as $\mathcal{O}(1/\sqrt{RD})$; thus we can set $R=1$ and retrieve sufficiently accurate results by taking large system sizes $D = 2^N$~\cite{Sum-24a}. This approach replaces the stochastic average with the expectation value evaluating using single `typical' state $\ket{\psi_{\rm rand}}$ (i.e., a single realization of $|\psi_r\rangle$). By absorption of the identities into the state vector $\ket{\psi_r}$, exploitation of the projector property $(\hat{P}^\uparrow_{n^\prime})^2 = \hat{P}^\uparrow_{n^\prime}$, the cyclicity of the trace, and by moving out of the Heisenberg picture, we arrive at the following form for the correlation function:
\begin{equation}\label{eqn:stochastic-correlator-projected}
    C_{n,n'}(t) \simeq \bra{\psi_{\rm rand}} \hat{P}^\uparrow_{n^\prime} \hat{\mathbb{U}}^{\dagger t} \hat{S}_n^z \hat{\mathbb{U}}^{ t} \hat{P}^\uparrow_{n^\prime} \ket{\psi_{\rm rand}}.
\end{equation}
We can interpret \Cref{eqn:stochastic-correlator-projected} as corresponding to an average of $\hat{S}_n^z$ expectation values computed over (un-normalized) trajectories $\ket{\overline{\psi_{n^\prime}}(t)} = \hat{\mathbb{U}}^{ t} \hat{P}^\uparrow_{n^\prime} \ket{\psi_{\rm rand}}$ of random initial states, which subsequently have their $n^\prime$-th site projected onto the spin-up state, and are then propagated in time by the repeated application of $t$ Floquet unitary layers. This approach has two main advantages: it gives us a clear prescription for computing correlation functions given limited resources, and it provides an intuitive connection between the significance that correlation functions have with respect to transport due to the revelation that they are composed of initial excitations which then spread. 

In this article, we restrict our analysis to systems with periodic boundary conditions, and - whilst individual circuits are not translationally invariant - we thus expect the transport properties to be translationally invariant given sufficient realizations. Thus without loss of generality we set $n^\prime = N/2$, and consider the magnetization profile:
\begin{equation}\label{eqn:spin-profile-projected}
    M_{n}(t) = \frac{1}{\Lambda} \bra{\overline{\psi_{N/2}}(t)} \hat{S}^z_n \ket{\overline{\psi_{N/2}}(t)} \approx \frac{1}{\Lambda} C_{n,N/2}(t)
\end{equation}
where $\Lambda = |\langle\overline{\psi_{N/2}}(0)\ket{\overline{\psi_{N/2}}(0)}|^2 \approx 1/2$ accounts for normalization of the initial state. Importantly, \Cref{eqn:spin-profile-projected} connects the magnetization profile to the correlation functions, which in turn encode information about transport. This approach offers a straightforward method for analyzing transport in large systems: generate a random state, project a single site onto the local spin-up state, dynamically evolve the state, and then probe the magnetization profile.

Due to our decision to interrogate systems with a $\mathrm U(1)$ symmetry, we further insist that the initial typical state $\ket{\psi_{\rm rand}}$ is drawn exclusively from a subsector of fixed magnetization $\langle \hat{S}^z_\text{tot}\rangle = M = \sum_n M_n(t)$. As discussed in the main text, we considered the zero-magnetization $M=0$ subsector throughout. This is the largest possible subsector in the total space, with dimension $\binom{N}{N/2}$, and thus yields the most accurate typical-state trace estimation (as per our discussion of \Cref{eqn:stochastic-trace-estimation}).

It is instructive to now determine the initial magnetization profile $M_n(0)$. The initial state $\frac{1}{\sqrt{\Lambda}}\hat{P}^\uparrow_{N/2}\ket{\psi_{\rm rand}}$ is prepared as per the discussion of \Cref{eqn:stochastic-correlator-projected} and \Cref{eqn:spin-profile-projected} and as discussed, for large systems it becomes a good approximation of \Cref{eq:initial-state}. 
%It bears a spin inhomogeneity $M_{N/2}(0) = 1/2$, embedded into a roughly homogeneous random background. If the random state is drawn from the subsector with total magnetization $M = \langle \hat{S}_\text{tot}^z \rangle$, then this homogeneous background consists of sites with local magnetization that fluctuates around $M_{n\neq N/2}(0) \simeq (M-1/2)/(N-1) = M_{\rm B}$ such that the total magnetization $M = \sum_{n}M_n(0)$ is conserved. 
%This behavior is captured in the following description of the magnetization profile,

% where $\text{d}W_n$ is some random process subject to the constraint $\sum_{n\neq N/2} \text{d}W_n = 0$ which accounts for random deviations in $|\psi_{\rm rand}\rangle$.

Finally, we emphasize that a single realization of our model consists of (a) drawing a random state $\ket{\psi_{\rm rand}}$ from which the initial state is generated, (b) generating spatial disorder embodied by drawing random phases $\{(\alpha_n, \beta_n, \phi_n)\}$, and (c) generating a random permutation $P(n)$ to define the circuit architecture. As per the typicality discussion above, the randomization over initial states is of negligible importance; we simply include it for completeness. The transport of the spin inhomogeneity at $n=N/2$, as captured by the magnetization profile $M_n(t)$, and the extraction of transport exponents are the foci of this study.

\subsection{Dynamical correlation function in the \textit{M}-subspace}
\label{supp:correlation_M}
The discussion can be adapted to our situation, where instead of working on the full Hilbert space, we limit it to $\mathcal{H}^{(M)}$. 
Let us define the dynamical correlation function restricted to the $M$-subspace as:
\begin{equation}
    C_{n,N/2}^{(M)}(t) = \Tr[\hat{P}_M\hat{S}_{N/2}^z(t) \hat{S}_{n}^z]
\end{equation}
where $\hat{P}_M=\sum_{i=0}^{d-1}\ket{i_M}\bra{i_M}$ is the projector onto the magnetization subspace (with the basis $\{\ket{i_M}\}_{i=0}^{d-1}$). 
We can check how $C_{n,N/2}^{(M)}(t)$ is related to $C_{n,N/2}(t)$.
Since $\Tr[\bullet]\equiv \sum_{M=-N/2}^{N/2}\Tr[\hat{P}_M\bullet]$ we have:
\begin{equation}
    C_{n,N/2}(t) = \sum_{M=-N/2}^{N/2} C_{n,N/2}^{(M)}(t).
\end{equation}
We can notice that:
\begin{equation}
\begin{split}
    \Tr[\hat{P}_M\hat{S}_n^z(t)] \stackrel{(1)}{=} & \Tr[\hat{P}_M \hat{S}_n^z]\\
     \stackrel{(2)}{=} & \frac{1}{N}\sum_{n=0}^{N-1}\Tr[\hat{P}_M\hat{S}_n^z]\\
    =& \frac{1}{N}\Tr[\hat{P}_M\hat{S}_\text{tot}^z] = \frac{M d_N^M}{N}
\end{split}
\end{equation}
where $(1)$: $[\hat{P}_M,\hat{\mathbb{U}}]=0$ and $(2)$: since $\hat{P}_M$ is invariant under permutation of spins, and $d_N^M=\binom{N}{\frac{N}{2}+M}$. 
Therefore
\begin{equation}
    C_{n,N/2}^{(M)}(t) = \Tr[\hat{P}_M\hat{S}_{n}^z(t) \hat{P}^\uparrow_{N/2}] - \Tr[\hat{P}_M\hat{S}_{n}^z(t)]/2
\end{equation}
becomes
\begin{equation}
    C_{n,N/2}^{(M)}(t) = \Tr[\hat{P}_M\hat{S}_{n}^z(t) \hat{P}^\uparrow_{N/2}] - \frac{M d_N^M}{2N}.
\end{equation}
We can now interpret $C_{n,N/2}^{(M)}(t)$ as the evolution of a perturbed initial state $\hat{\rho}(0)$, followed by a measurement, similarly as done for $C_{n,N/2}(t)$.
We have defined the initial state as:
\begin{equation}
    \hat{\rho}(0) = \hat{P}_M \hat{P}^\uparrow_{N/2} \hat{P}_M
    \tag{\ref{eq:initial-state}}
\end{equation} 
and since $[\hat{P}^\uparrow_{N/2}, \hat{P}_M] = 0$ this leads to:
\begin{equation}
\begin{split}
    C_{n,N/2}^{(M)}(t) = \Tr[\hat{S}^z_n(t)\hat{\rho}(0)] - \frac{M d_N^M}{2N}\,.
\label{eq:correlation_to_magnetization}
\end{split}
\end{equation}

\subsection{Dynamical correlation function in the \textit{M}-subspace through typicality}
\label{supp:dynamical_correlation}

The last step remaining to connect $M_n(t)$ with the correlation functions involves typical states in the $\mathcal{H}^{(M)}$ space. In Ref.~\cite{Ric-21a}, it was shown how to calculate the (full space) dynamical correlation function $C_{n,N/2}(t)=\Tr[\hat{S}_n^z(t)\hat{S}_{N/2}^z(t)]$ by measuring an expectation value of a time-evolved typical random state. The basic idea is to use the fact that for a typical random state $\ket{\psi_{\rm rand}}$, in a $d$-dimensional Hilbert space, we have:
\begin{equation}
    \bra{\psi_{\rm rand}}\hat{A} \ket{\psi_{\rm rand}} \simeq \frac{1}{d}\Tr[\hat{A}]+O\left(\frac{1}{d}\right)
\end{equation}
for any observable $A$ (see also Ref.~\cite{Wei-06a}), i.e., the trace of an operator is well approximated by taking its expectation value with respect to $\ket{\psi_{\rm rand}}$. This is useful for computing dynamical correlation functions of \Cref{eq:correlation_to_magnetization}. Such a trace can be computed via typical random states by:
\begin{equation}
\begin{split}
    \Tr[\hat{S}^z_n(t)\hat{\rho}(0)] &= \Tr[\hat{P}_M \hat{S}^z_n(t) \hat{P}_M \hat{P}^\uparrow_{N/2}  ] \\
    &\simeq \frac{d}{\tilde{d}} \bra{\psi_{\rm rand}} \hat{P}_M \hat{P}^\uparrow_{N/2} \hat{S}_n^z(t) \hat{P}^\uparrow_{N/2} \hat{P}_M \ket{\psi_{\rm rand}}\\
    &\simeq \frac{d}{\tilde{d}} \bra{\psi^{(M)}_{\rm rand}} \hat{P}^\uparrow_{N/2} \hat{S}_n^z(t) \hat{P}^\uparrow_{N/2} \ket{\psi^{(M)}_{\rm rand}}\\
    &\simeq \frac{d}{\tilde{d}} \bra{\overline{\psi_{n^\prime}}(t)}\hat{S}_n^z\ket{\overline{\psi_{n^\prime}}(t)}.
\end{split}
\end{equation}
So eventually, measuring the magnetization of $\ket{\overline{\psi_{n^\prime}}(t)}$ corresponds to measure the correlation function as
\begin{equation}
    C_{n,N/2}^{(M)}(t) \simeq \frac{d}{\tilde{d}} \bra{\overline{\psi_{n^\prime}}(t)}\hat{S}_n^z\ket{\overline{\psi_{n^\prime}}(t)} - \frac{M d_N^M}{2N}.
\end{equation}

% markov chain:

Finally, (ii) we suggest a potential mechanism for the prethermal behaviour in the swappy regime. We suggest that it is due to the ``fragmentation'' of the initial excitation into local populations reminiscent of the on-site populations of particles undergoing (potentially directionally-biased) one-dimensional random walks; i.e. that the swappy regime is an essentially single-particle phenomenon. Our proposed mechanism is as follows: near the SWAP point, the effect of local unitaries $\hat{U}_{n,n+1}$ on the local magnetization $s^z_n$ on site $n$ is realized in two ways, (i) the exchange of most ($p_n$) of $s^z_n$ to the $n+1$th site, and (ii) a remnant ($1-p_n$) of $s^z_n$ that is left behind on the original site $n$. Essentially, we can approximate the dynamics close to the SWAP point as a classical Markov chain realizing the evolution of populations according to a classical update rule such as: $s^z_{n+1} = (1-p_{n+1})s^z_{n+1} + p_{n}s^z_n$. For imperfect SWAP gates, $p_n < 1$, excitations recursively ``split'' as consecutive imperfect SWAP gates are applied. We can explicitly see this splitting behavior realized as a `checkerboard' pattern in \Cref{fig:magnetization_profiles}\subfig{c} and (to a lesser extent due to averaging) \subfig{g}. As $J \to \pi$ this splitting behavior vanishes, as is seen in the near-SWAP regime of \Cref{fig:magnetization_profiles}\subfig{d} and \subfig{h}, and we retrieve perfect swap dynamics, even at late times. This proposed mechanism is exact for small perturbations $J^\prime$ away from the SWAP point along the diagonal $J=J_z=\pi-J^\prime$, wherein (neglecting the local disorder terms, and global phases) the local unitary can be decomposed as:
\begin{align}
    \hat{U}_{n,n+1} &= e^{-i(\pi-J^\prime) \hat{\mathbf{S}}_n \cdot \hat{\mathbf{S}}_{n+1}} \\ 
    &= \swap_{n,n+1}\left[\mathbb{I} + i J^\prime \hat{\mathbf{S}}_n \cdot \hat{\mathbf{S}}_{n+1}\right] + \mathcal{O}(J^{\prime 2}) \\
    &= \left(1-\frac{iJ^\prime}{4}\right)\swap_{n,n+1} + i \frac{J^\prime}{2} \mathbb{I} + \mathcal{O}(J^{\prime 2}) \label{eqn:jprime-perturbed}
\end{align}
such that, after application to some test state,
\begin{equation} 
    \hat{U}_{n,n+1} |\psi\rangle \approx \frac{1}{\sqrt{\mathcal{N}}} \left[ \left(1-\frac{iJ^\prime}{4}\right) |\widetilde{\psi}\rangle + i \frac{J^\prime}{2} |\psi\rangle  \right]
\end{equation}
where $\swap|\psi\rangle = |\widetilde{\psi}\rangle$, and $\mathcal{N}$ is some appropriate normalization constant. This brings the Markov chain description of short-time dynamics close to the SWAP point $J^\prime = 0$ discussed above into sharp clarity; with the probabilities $p_n$ determined by the value of $J^\prime$. As $J^\prime$ increases, the $\mathcal{O}(J^\prime)$ term in \Cref{eqn:jprime-perturbed} yields noticeable effects on the dynamics at earlier times. We interrogate these effects here by extracting prethermal timescales, where we find numerical evidence supporting our above discussion. However, a detailed semi-classical analysis of the swappy regime according to this intuition, and its connection to the sudden thermalization seen in \Cref{fig:magnetization_profiles}\subfig{c} and \subfig{g}, is beyond the scope of this paper, and is deferred to future study.

\end{document}